\documentclass[a4paper,10pt]{article}

\usepackage[T1]{fontenc}
\usepackage{amsmath,amsfonts,amssymb,amsthm,bbm}
\usepackage{graphicx}
\usepackage{color}
\usepackage{comment}
\usepackage{pifont}

\allowdisplaybreaks

\usepackage[english]{babel}

\newtheorem{definition}{Definition}

\newtheorem{innercustomdef}{Definition}
\newenvironment{custom.definition}[1]
  {\renewcommand\theinnercustomdef{#1}\innercustomdef}
  {\endinnercustomdef}

\newtheorem{part-ent}{Particle-entanglement}
\newtheorem*{mode-ent*}{Mode-locality and mode-entanglement}

\newcommand{\cmark}{\ding{51}}
\newcommand{\xmark}{\ding{55}}

\title{Entanglement in indistinguishable\\ particle systems}

\author{F. Benatti$^a$, R. Floreanini$^b$, F. Franchini$^c$, U. Marzolino$^{b,c}$ \\
\\
\small ${}^a$Dipartimento di Fisica, Universit\`a di Trieste, I-34151 Trieste, Italy \\
\small ${}^b$Istituto Nazionale di Fisica Nucleare, Sezione di Trieste, I-34151 Trieste, Italy \\
\small ${}^c$Institut Ru\d er Bo\v{s}kovi\'c, HR-10000 Zagreb, Hrvatska}

\date{}

\begin{document}

\maketitle

 




%
%
%
%
%
%
%

\begin{abstract}
For systems consisting of distinguishable particles, there exists an agreed upon notion of  entanglement 
which is fundamentally based  on the possibility of addressing individually each one of the constituent parties.
Instead, the indistinguishability of identical particles hinders their individual addressability and has prompted diverse,
sometimes discordant definitions of entanglement. 
In the present review, we provide a comparative analysis of the relevant existing approaches, which is based on the characterization of bipartite
entanglement in terms of the behaviour of correlation functions. Such a point of view provides a fairly general setting where to discuss the presence of non-local effects; it is performed in the light
of the following general consistency criteria:
\textit{i)} entanglement corresponds to non-local correlations and cannot be generated by local operations; \textit{ii)} when, by ``freezing'' suitable degrees of freedom, identical particles can be effectively distinguished, their entanglement must reduce to
the one that holds for distinguishable particles; \textit{iii)} in absence of other quantum resources, only entanglement can
outperform classical information protocols. 
These three requests provide a setting 
that  allows to evaluate strengths and weaknesses of the existing approaches to
indistinguishable particle entanglement and to contribute to the current understanding of such a crucial issue.
Indeed, they can be classified into five different classes: four hinging on the notion of particle and one based on
that of physical modes. We show that only the latter approach is consistent with all three criteria, each of the others indeed violating at least one of them.
\end{abstract}



%
%
%

\section{Introduction}
\label{Intro}

Entanglement is a peculiar feature of compound quantum systems that is rooted in the superposition principle. While it
challenges our common sense notion of locality~\cite{BellGao},
it also represents a practical resource for quantum technologies that will enable them to outperform classical devices, 
thus heralding a new technological revolution~\cite{NielsenChuang,BenattiFannesFloreaniniPetritis,
Buhrman2010,BrussLeuchs}. This latter motivation has boosted plenty of experimental efforts and theoretical proposals  aiming at generating entanglement and at applying it to practical tasks in information transmission and manipulation.

Considerable efforts have been devoted to the study of entanglement in compound systems consisting of distinguishable particles~\cite{Amico2008,Horodecki2009,Guhne2009,
BenattiFannesFloreaniniPetritis,BrussLeuchs}, e.g. spin models. Indeed, quantum devices hinging on the notion of qubit require that particles be addressed and manipulated individually.
Particle distinguishability is assumed also for $d$-level systems \cite{Fickler2012,Malik2016} and continuous variables \cite{Braunstein2005}.
For the aforementioned systems, there are several criteria to detect and quantify entanglement, e.g. the partial transposition criterion \cite{Peres1996,Horodecki1996}, the criterion derived in \cite{Duan2000}, the Simon criterion \cite{Simon2000}, the Hillery-Zubairy inequalities \cite{Hillery2006}, the van Loock-Furusawa inequalities \cite{vanLoock2003}, entanglement monotones \cite{Vidal2000,Plenio2005}, and entanglement witnesses \cite{Terhal2000,Chruscinski2014}.
Furthermore, entanglement is a powerful tool that has been used to characterize several phenomena in many-body systems \cite{Amico2008,Tichy2011}, like phase transitions \cite{Sachdev2011}, area laws \cite{Eisert2010}, and many-body localization \cite{Abanin2019}.

In the case of bipartite systems consisting of  two particles $S_1$ and $S_2$,  their individual addressability leads to
identify entangled state vectors of $S_1+S_2$ as those that are not of the so-called separable tensor product form 
$\vert\psi^{\textnormal{sep}}_{12}\rangle=\vert\psi_1\rangle\otimes\vert\psi_2\rangle$, where $\vert\psi_1\rangle$ is a state vector of $S_1$ and $\vert\psi_2\rangle$ of $S_2$.
Separable state vectors $\vert\psi^{\textnormal{sep}}_{12}\rangle$ can also be equivalently identified as those states that do not support correlations between 
any pair of observables $A$ of $S_1$  and $B$ of $S_2$:
\begin{equation}
\label{fact}
\langle\psi^{\textnormal{sep}}_{12}\vert A\,B\vert\psi^{\textnormal{sep}}_{12}\rangle\,=\,\langle\psi^{\textnormal{sep}}_{12}\vert A\vert\psi^{\textnormal{sep}}_{12}\rangle\,\langle\psi^{\textnormal{sep}}_{12}\vert B\vert\psi^{\textnormal{sep}}_{12}\rangle\ .
\end{equation}

As the observables $A$ and $B$ refer to different degrees of freedom of particle type, they commute: $[A,B]=0$. They can thus be dubbed local from the particle point of view, a fact expressed by 
writing them as $A\otimes \mathbbm{1}$ and $\mathbbm{1}\otimes B$, \textit{i.e.} as observables of the two-particle system $S_1+S_2$. As a consequence, the factorization in~\eqref{fact} when holding for all particle-local $A$ and $B$ also expresses the absence of correlations that would be non-local from the particle point of view.
Instead, entangled state vectors $\vert\psi^{\textnormal{ent}}_{12}\rangle$ of $S_1+S_2$ carry statistical correlations between
at least one pair of commuting,  particle-local observables. These correlations are identified by the lack of factorization of the corresponding two-point correlation functions
\begin{equation}
\label{non-fact}
\langle\psi^{\textnormal{ent}}_{12}\vert A\,B\vert\psi^{\textnormal{ent}}_{12}\rangle\ne\langle\psi^{\textnormal{ent}}_{12}\vert A\vert\psi^{\textnormal{ent}}_{12}\rangle\,\langle\psi^{\textnormal{ent}}_{12}\vert B\vert\psi^{\textnormal{ent}}_{12}\rangle\ ,
\end{equation} 
and are thus non-local from the particle point of view.

In the case of indistinguishable particles, the fact that  individual constituents cannot be distinguished has two consequences: under particle exchange, 
bipartite state vectors must be symmetric (Bosons)  or anti-symmetric (Fermions) and single particle observables must be symmetric~\cite{Landau,MessiahII,Feynman}.
Therefore, neither particle-separable states as $\vert\psi^{\textnormal{sep}}_{12}\rangle$, with $\vert\psi_1\rangle\neq\vert\psi_2\rangle$  are \textit{bona fidae} state vectors for two identical particles nor 
$A\otimes\mathbbm{1}$ and $\mathbbm{1}\otimes B$  are physically acceptable as particle-local observables.

In the following we shall address various answers to the following two questions that have been proposed in the years: 
\begin{itemize}
\item Are states of two identical particles of the from $\vert\psi_1\rangle\otimes\vert\psi_2\rangle\pm\vert\psi_2\rangle\otimes\vert\psi_1\rangle$ really entangled as their
form would suggest? 
\item Which is the idea of locality that underlines the answer to the previous question? 
\end{itemize}

In addressing these two issues, we will shift the  focus
from the analysis  of the tensor product structure of quantum states to the properties of specific, commuting subsets of local observables able to expose
the lack of factorization of two-point correlation functions~\cite{BratteliRobinson,Strocchi}.
In this respect, we shall take the notion of locality of quantum observables as the primary notion to consider when deciding which states are entangled and which ones are separable~\cite{Werner1989}. Entanglement is in fact synonym of non-local correlations which however require a locality criterion in order to be defined in a context without an a priori given natural tensor product structure. Indeed,  more or less implicitly, each approach to identical particle entanglement unavoidably entails an underlying locality criterion of its own.

The factorization of two-point correlation functions, or lack of it thereof, is a clear evidence of the absence or presence of correlations 
both for identical and for distinguishable particles. 
While it is known that this is not the case for genuine multipartite systems consisting of distinguishable particles~\cite{Kaszlikowski2008,Bennett2011,Schwemmer2015,Tran2017,
Designolle2017} requiring a subtler definition of locality, for identical particles a consistent theory of multipartite entanglement has not yet been fully developed~\cite{BrussLeuchs}. In this review we restrict to generalized bipartite settings where two point correlation functions provide  a universal tool to identify the classes  of separable and entangled quantum states. In full generality, however, the properties shared by states in different classes dramatically depend on which observables are considered local and which are not.

The essential dependence of entanglement and separability on
the choice of local and non-local observables, which will result to be prominent for identical particles, can however be illustrated already in the case of two distinguishable qubits. In fact, let us consider the Bell states 
\begin{equation}
\label{Bell1}
|\Psi_\pm\rangle=\frac{\vert\!\uparrow\uparrow\rangle\pm\vert\! \downarrow\downarrow\rangle}{\sqrt{2}}\ ,
\end{equation}
where  $\vert\!\uparrow\rangle,\vert\!\downarrow\rangle$ are the eigenstates of the Pauli matrix $\sigma_3$. These states are prototypical entangled states from the particle point of view: 
the non-local correlations supported, for instance, by $\vert\Psi_+\rangle$,  are witnessed by lack of factorization in the expectations of products of suitable commuting single-particle observables like $P_1=\vert\!\uparrow\rangle\langle\uparrow\!\vert\otimes \mathbbm{1}$ and $P_2=\mathbbm{1}\otimes\vert\!\downarrow\rangle\langle\downarrow\!\vert$, that are local from the particle point of view. Indeed, one finds
\begin{equation}
\label{Bell2}
\langle\Psi_+\vert P_1P_2\vert\Psi_+\rangle=\frac{1}{2}\neq \frac{1}{4}=\langle\Psi_+\vert P_1\vert\Psi_+\rangle\,\langle\Psi_+\vert P_2\vert\Psi_+\rangle\ .
\end{equation}
On the other hand, one can construct commuting
observables that do not correspond to single-particle properties; as an example, consider the
observables of the form
$\mathcal{O}_\pm=\alpha_\pm\,\mathbbm{1}+\beta_\pm\,\vert\Psi_\pm\rangle\langle\Psi_\pm\vert$ with $\alpha_\pm$ and $\beta_\pm$ real constants.
They commute, but correspond to global properties shared by both particles, whence they are not particle-local; on the other hand, one checks that the state $\vert\Psi_+\rangle$ is separable with respect to
the non-particle local observables $\mathcal{O}_+$ and $\mathcal{O}_-$; indeed,
\begin{align}
\label{Bell3}
\langle\Psi_+\vert \mathcal{O}_+\mathcal{O}_-\vert\Psi_+\rangle & =\langle\Psi_+\vert \mathcal{O}_+\vert\Psi_+\rangle\,\langle\Psi_+\vert\mathcal{O}_-\vert\Psi_+\rangle \nonumber \\
& =\alpha_-(\alpha_++\beta_+)\ .
\end{align}
Then, the state $\vert\Psi_+\rangle$ is entangled with respect to the locality criterion based on the particle point of view, but separable relative to the locality criterion based on the commuting
observables $\mathcal{O}_\pm$~\cite{Harshman2011,Thirring2011}. 
 
In recent years, several
definitions of entanglement of identical particles have been proposed with different consequences on their usability in information protocols, resulting in a confusing plethora of more or less discordant possibilities and still continuing debates:
we will refer to the relevant literature while expounding the various approaches.
Here it suffices to say that the approaches to identical particle entanglement can be divided into five classes. Three of them explicitly  focus upon correlations among particles and use the first quantization formalism: these ones will be listed under the caption entanglement-I, -II, -III. A fourth one, though using the second quantization formalism, does nevertheless still refer to the notion of particle and will be denoted as entanglement-$IV$. The last class instead considers correlations among the possible modes available to the identical particles which, unlike particles, are always singly addressable~\cite{Wurtz2009,Bakr2010,Sherson2010}. This approach makes use of the second quantization formalism and will be denoted as entanglement-V.

In addition, a point should be stressed here: indistinguishable particles can be effectively distinguished by ``freezing'' some of their degrees of freedom, for instance by confining particles within finite spatial regions: these ``frozen'' degrees of freedom can then be used to unambiguously label each particle \cite{DeMuynck1975,Herbut1987,Herbut2001,Herbut2006,
Tichy2013,Cunden2014},
as proposed in several condensed matter implementations~\cite{Aspect1981,Ionicioiu2002,Negretti2011,Rohling2012,Underwood2012,Inaba2012}.
Then, the entanglement properties of indistinguishable particles made distinguishable by ``freezing'' suitable degrees of freedom, should reduce to the standard ones.

As identical particles are at the root of many-body systems and since these latter are the building blocks of most new generations of
quantum devices, providing for them 
a physically sensible
and practically useful
notion of entanglement is therefore of utmost importance. This is especially true for integrated architectures where the degrees of freedom involved in implementing a certain task and allowing for device scalability cannot in principle
be used to label particles.
Progress in this direction have been done for ultracold atoms \cite{Underwood2012,Inaba2012}, quantum optics \cite{Pan2012,Roslund2013,Chen2014,Takeda2019,Korolkova2019,
Fabre2019,Sandbo2020}, and quantum fields \cite{Narnhofer2002,Casini2009,Calabrese2009,Calabrese2016,
Rangamani2017,HollandSanders,Nishioka2018,Witten2018}.

In this review, we refer to non-locality as defined by non-vanishing
correlation functions of commuting observables in a bipartite scenario. In all cases, we do not refer to non-locality as related
to the violations of Bell's inequalities. We also discuss entangled
resources identified by these bipartite correlations and leave aside
general resource theories \cite{Chitambar2019,ContrerasTejada2019,Chitambar2020}. Our aim is the study of their compatibility with respect to 
three very general physical requirements that are necessary in order to guarantee the full 
consistency of entanglement theory, both for distinguishable and identical particles:

\begin{enumerate}
\item {\bf Local operators.}
As for distinguishable particles, entanglement for identical particles must correspond to the presence of non-local correlations, with respect to
an appropriate definition of locality.
Given a locality criterion, observables that are considered local must satisfy the factorization properties \eqref{fact} and \eqref{non-fact} for separable and entangled states respectively, and cannot generate entanglement.
\item {\bf Effective distinguishability.} The entanglement theory developed for identical particles must reduce to the standard one when applied to identical particles that have effectively
become distinguishable by ``freezing'' suitable degrees of freedom.

\item {\bf Information processing resources.} In absence of other quantum resources,
local operators acting on separable states must not enhance the performances of informational tasks with respect to classical ones.
\end{enumerate}

We shall compare the existing approaches to identical particle entanglement by means of the three above criteria and show that entanglement-I, -II, -III and -IV violate at least one criterion, while entanglement-V fulfils all of them. More in detail, in Section~\ref{eff.dist}, we briefly review the formalism necessary to tackle systems consisting of identical particles and the techniques useful to effectively distinguish them.
In Section~\ref{part-ent} we shall focus upon various notions of particle entanglement formulated within the first quantization approach to many-body quantum systems that have appeared in the literature;  
we regroup them into the four categories of entanglement-I, -II , -III and -IV, discussing their strengths and weaknesses with respect to  the above three physical criteria.
Section~\ref{mode-ent} is instead devoted to the presentation of entanglement-V which is based on the second quantization approach to quantum many-body systems and focuses upon its compatibility with the chosen physical criteria. In the final Section we summarize the outcomes of our analysis,
pointing to their relevance  in quantum technologies, while some more technical details are collected in two Appendices.

\section{Identical and distinguishable particles} \label{eff.dist}

Quantum mechanics states that the wave-functions of identical particles must be symmetric or antisymmetric under the exchange of particle labels~\cite{Landau,MessiahII,Feynman}.
Nevertheless, even identical particles,
for instance when they are sufficiently far apart from each other, may be considered to behave as if they were distinguishable,
and may thus be labelled by incompatible values of suitable observables that remain constant in time~\cite{DeMuynck1975,Herbut1987,Herbut2001,Herbut2006,Tichy2013,Cunden2014}.
\medskip

The physical idea behind
the notion of effective distinguishability is that observables that might expose
particle identity are difficult to be accessed in practice. This is the case,  for instance, when one tries to determine whether there are spatially non-local correlations among particles far away from each other. Any degree of freedom can be used to label identical particles, for instance momentum or energy eigenstates, or other internal degrees of freedom, such as spin. For the sake of simplicity, in the following
we will effectively distinguish identical particles by localizing them within non-overlapping spatial regions. 

In the case of two distinguishable particles, labelled by $1$ and $2$, their total Hilbert space, $\mathcal{H}_{12}$, is the tensor product, $\mathcal{H}_{12}:=\mathcal{H}_1 \otimes \mathcal{H}_2$, of the single-particle Hilbert spaces.
Furthermore, single-particle operators act on $\mathcal{H}_{12}$ as $O_1\otimes\mathbbm{1}$ and $\mathbbm{1}\otimes O_2$, with $\mathbbm{1}$ the identity operator and $O_j$ any operator
acting on $\mathcal{H}_j$.
For two identical particles, instead, indistinguishability requires
their Hilbert space
to be the Bosonic symmetrization (Fermionic anti-symmetrization) of the tensor product $\mathcal{H}\otimes\mathcal{H}$ of a same single-particle Hilbert space $\mathcal{H}$. This is obtained by means of the symmetrization, $\mathcal{S}=\mathcal{S}^2$, and anti-symmetrization, $\mathcal{A}=\mathcal{A}^2$, projectors such that 
\begin{eqnarray}
\label{symproj}
\hskip-.5cm
\mathcal{S} \, \vert\psi\rangle\otimes|\phi\rangle & = & \frac{1}{2}\Big(\vert\psi\rangle\otimes|\phi\rangle+\vert\phi\rangle\otimes|\psi\rangle\Big), \\
\label{asymproj}
\hskip-.5cm\mathcal{A} \, \vert\psi\rangle\otimes|\phi\rangle & = & \frac{1}{2}\Big(\vert\psi\rangle\otimes|\phi\rangle-\vert\phi\rangle\otimes|\psi\rangle\Big)\ ,
\end{eqnarray}
for all $|\psi\rangle$ and $|\phi\rangle$ in the single-particle Hilbert space  $\mathcal{H}$.

Let the symbol $\mathfrak{S}$ stand for either $\mathcal{S}$ or $\mathcal{A}$, depending on the particle statistics.
Since $O_1\otimes\mathbbm{1}$ and $\mathbbm{1}\otimes O_2$ address individual particles, indistinguishability forces one to distribute the action of single-particle operators over all particles by means of the symmetrized single-particle operator
\begin{equation}
\label{idensingop}
\mathcal{P}(O,\mathbbm{1})=\mathcal{P}(\mathbbm{1},O)=O\otimes\mathbbm{1}+\mathbbm{1}\otimes O\ .
\end{equation}
In the case of two-particle operators, particle indistinguishability demands them to be of the form 
\begin{equation}
\label{loc.op}
\mathcal{P}(O_1,O_2) \equiv O_1 \otimes O_2 + O_2 \otimes O_1, 
\end{equation}
that can be recast as 
\begin{equation}
\label{loc.op2} 
\mathcal{P}(O_1,O_2)= 2\, \mathcal{S}(O_1 \otimes O_2) \mathcal{S} + 2 \,\mathcal{A}(O_1 \otimes O_2)\mathcal{A}\ .
\end{equation}

In order to provide a convenient reference framework for what follows, we will consider particles described by an ``external'' degree of freedom with values $L,R$, and an ``internal'' degree of freedom with values $\uparrow,\downarrow$.
The single-particle Hilbert space  is then spanned by the four orthogonal states $\vert L,\uparrow\rangle$, $\vert L,\downarrow\rangle$, $\vert R,\uparrow\rangle$ and $\vert R,\downarrow\rangle$. Among the vectors of the two-particle (anti-)symmetrized Hilbert space we will in particular consider those of the form
\begin{align}
\nonumber
\vert\phi^\textnormal{id}_1\rangle
  & =\sqrt{2}\,\mathfrak{S} \Big[  \vert L,\uparrow\rangle\otimes\vert R,\downarrow\rangle \Big]\\
\label{eff.dist.sep}
  & =\frac{1}{\sqrt{2}}\Big[  \vert L,\uparrow\rangle\otimes\vert R,\downarrow\rangle \pm
  \vert R,\downarrow\rangle\otimes\vert L,\uparrow\rangle \Big]\ , \\
\nonumber
\vert\phi^\textnormal{id}_2\rangle
  & = \mathfrak{S} \Big[
  \vert L,\uparrow\rangle\otimes\vert R,\downarrow\rangle+\vert L,\downarrow\rangle\otimes\vert R,\uparrow\rangle\Big]\\
\nonumber
& =\frac{1}{\sqrt{2}}\Big[  \vert L,\uparrow\rangle\otimes\vert R,\downarrow\rangle +\vert L,\downarrow\rangle\otimes\vert R,\uparrow\rangle \\
&\pm \vert R,\downarrow\rangle\otimes\vert L,\uparrow\rangle \pm
  \vert R,\uparrow\rangle\otimes\vert L,\downarrow\rangle \Big]\ .
  \label{eff.dist.ent}   
\end{align}
Such a representation of identical two-particle states is
typical of the first quantization approach; a much more effective technique to deal with any number of identical particles, as in many-body theories, is provided by second quantization in terms of annihilation and creation operators $\mathfrak{a}_f\,,\,\mathfrak{a}_f^\dag$, where $\mathfrak{a}_f^\dag$
creates the single-particle vector state $\vert f\rangle$  by acting on the selected vacuum vector, $\vert\textnormal{vac}\rangle$, of the theory, $\mathfrak{a}_f^\dag\vert\textnormal{vac}\rangle=\vert f\rangle$, while $\mathfrak{a}_f\vert\textnormal{vac}\rangle=0$.
In the second quantization formalism, the states~\eqref{eff.dist.sep} and~\eqref{eff.dist.ent}  then read
\begin{align}
\label{eff.dist.sep2}
\vert\phi^\textnormal{id}_1\rangle
  &  =\mathfrak{a}_{L,\uparrow}^\dag\mathfrak{a}_{R,\downarrow}^\dag\vert\textnormal{vac}\rangle, \\
\label{eff.dist.ent2}
\vert\phi^\textnormal{id}_2\rangle
  & =\frac{1}{\sqrt{2}}\left(\mathfrak{a}_{L,\uparrow}^\dag\mathfrak{a}_{R,\downarrow}^\dag+\mathfrak{a}_{L,\downarrow}^\dag\mathfrak{a}_{R,\uparrow}^\dag\right)\vert\textnormal{vac}\rangle\ ,
\end{align} 
where $\mathfrak{a}_{S,\sigma}^\dag$ is the creation operator of a particle with a state labelled by $S\in\{L,R\}$ and $\sigma\in\{\uparrow,\downarrow\}$:
\begin{equation}
\label{auxx2}
\mathfrak{a}^\dag_{S,\sigma}\vert\textnormal{vac}\rangle=\vert S,\sigma\rangle \ .
\end{equation}
Notice that, unlike in the first quantization formalism, now the necessary \hbox{(anti-)} symmetry of state vectors automatically follows from the (anti-)commutation relations:
\begin{equation}
\label{CCAR}
\Big[\mathfrak{a}_{S_1,\sigma_1}\,,\,\mathfrak{a}^\dag_{S_2,\sigma_2}\Big]_{\pm}=\delta_{S_1S_2}\,
\delta_{\sigma_1\sigma_2}\ ,
\end{equation}
where $\pm$ denote anti-commutator, respectively commutator.

\begin{figure}[htbp]
\centering
\includegraphics[width=\textwidth]{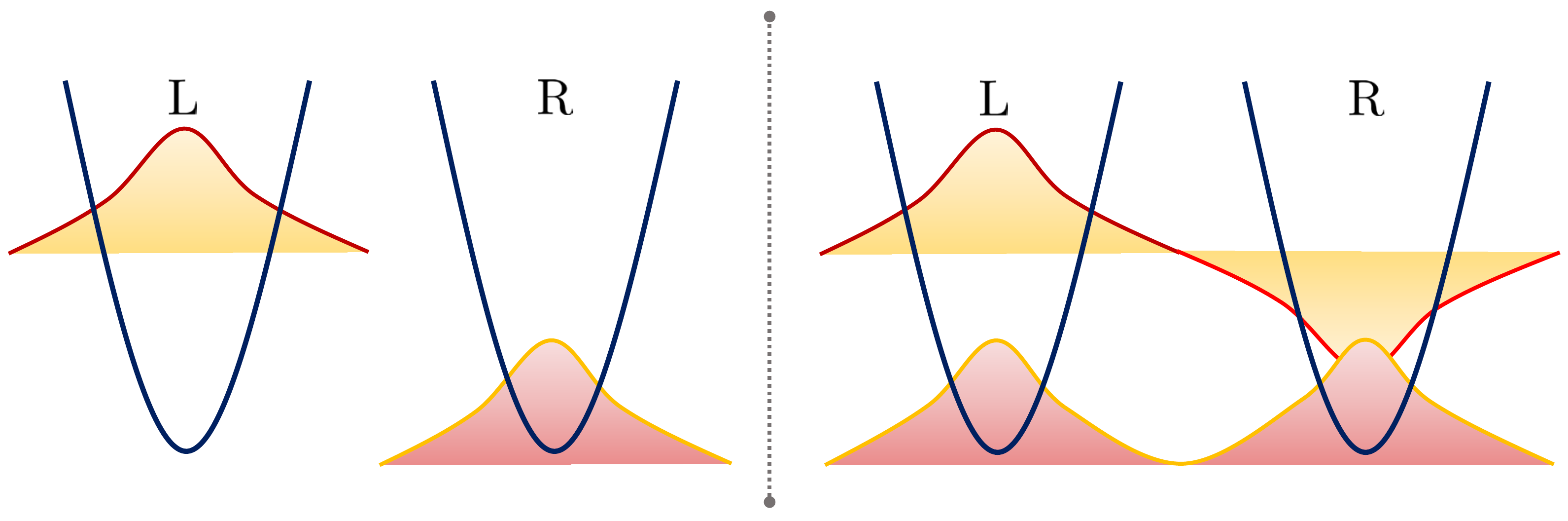}
\caption{Orthogonal spatial wave functions that can be used to effectively distinguish particles through  spatially localized (left panel) or spatially delocalized modes (right panel).}
\label{fig.dist}
\end{figure}

We now
briefly discuss how identical particles can be effectively distinguished (see \cite{DeMuynck1975,Herbut1987,Herbut2001,Herbut2006,Tichy2013,Cunden2014} for a more comprehensive presentation).
If $L$ and $R$
label orthogonal single-particle states,
as when particles belong to disjoint spatial regions, left and right, so that  $\langle L\vert R\rangle=0$, the external degree of freedom, identifiable as position, can be used to give particles a label, thus
effectively identifying them (see Figure \ref{fig.dist}). Indeed, by measuring and then ``freezing'' the spatial degree of freedom, one can label the particle found in the region $L$ as particle $1$, and the one found in $R$, as particle $2$. 

Of course, the label attribution is incompatible with particle identity; however,
suppose only observables of the form $\mathcal{P}(P_L\otimes\Sigma_1,P_R\otimes\Sigma_2)$ are experimentally accessible, where  
$P_{L,R}$ are projectors $\vert L\rangle\langle L\vert$ and $\vert R\rangle\langle R\vert$ onto states localized in the left, respectively right region, while $\Sigma_1$ and $\Sigma_2$ are arbitrary observables relative to the internal degree of freedom. Then, by measuring solely these observables, one cannot distinguish the states~\eqref{eff.dist.sep} 
and~\eqref{eff.dist.ent} from 
\begin{align}
\label{eff.dist.sep.isom}
\hskip-.2cm
|\phi^\textnormal{dist}_1\rangle & =|L,\uparrow\rangle\otimes|R,\downarrow\rangle\ ,\\
\label{eff.dist.ent.isom}
\hskip-.2cm
|\phi^\textnormal{dist}_2\rangle & =\frac{1}{\sqrt{2}}\big(|L,\uparrow\rangle\otimes|R,\downarrow\rangle+|L,\downarrow\rangle\otimes|R,\uparrow\rangle\big)\ ,
\end{align}
where the first particle is left localized and the second one right localized, so that
space localization allows one to give identical particle a label.
Indeed, exploiting spatial orthonormality,
the mean values
of the restricted set of observables $\mathcal{P}(P_L\otimes\Sigma_1,P_R\otimes\Sigma_2)$ read
\begin{align}
& \langle\phi^\textnormal{id}_1\vert\mathcal{P}\big(P_L\otimes\Sigma_1,P_R\otimes\Sigma_2\big)\vert\phi^\textnormal{id}_1\rangle= \nonumber \\
& =\langle L,\uparrow|P_L\otimes\Sigma_1|L,\uparrow\rangle\langle R,\downarrow|P_R\otimes\Sigma_2|R,\downarrow\rangle\nonumber\\
& =\langle\phi^\textnormal{dist}_1\vert \big(P_L\otimes\Sigma_1\big)\otimes \big(P_R\otimes\Sigma_2\big)\vert\phi^\textnormal{dist}_1\rangle\ ,
\label{effdist}
\end{align}
and similarly
\begin{align}
& \langle\phi^\textnormal{id}_2\vert\mathcal{P}\big(P_L\otimes\Sigma_1,P_R\otimes\Sigma_2\big)\vert\phi^\textnormal{id}_2\rangle= \nonumber \\
& =\langle\phi^\textnormal{dist}_2\vert \big(P_L\otimes\Sigma_1\big)\otimes \big(P_R\otimes\Sigma_2\big)\vert\phi^\textnormal{dist}_2\rangle\ .
\label{effdist2}
\end{align}
It is important at this point to make the following observations: firstly, observables as
\begin{equation}
\label{locpartobs}
\big(P_L\otimes\Sigma_1\big)\otimes \big(P_R\otimes\Sigma_2\big)\ ,
\end{equation}
are \textit{particle-local} since they individually address single-particles through the observables $\big(P_L\otimes\Sigma_1\big)\otimes \mathbbm{1}$ and 
$\mathbbm{1}\otimes\big(P_R\otimes\Sigma_2\big)$, while observables of the form $\mathcal{P}(P_L\otimes\Sigma_1,P_R\otimes\Sigma_2)$
are not particle-local. Secondly, the equalities 
in~\eqref{effdist} and \eqref{effdist2} show that such particle-non-locality
cannot be
witnessed by state vectors  as those in~\eqref{eff.dist.sep.isom} and~\eqref{eff.dist.ent.isom}.

Notice that $\mathcal{P}(P_L\otimes\Sigma_1,P_R\otimes\Sigma_2)$ reads $A_L A_R$  in second quantization, with
\begin{align}
\label{AL}
A_L & =\sum_{\sigma_1,\sigma_1'}\Sigma_1(\sigma_1,\sigma_1')\,\mathfrak{a}_{L,\sigma_1}^\dag\,\mathfrak{a}_{L,\sigma_1'}\ , \\
\label{AR}
A_R & =\sum_{\sigma_2,\sigma_2'}\Sigma_2(\sigma_2,\sigma_2')\,\mathfrak{a}_{R,\sigma_2}^\dag\,\mathfrak{a}_{R,\sigma_2'}\ ,
\end{align}
where $\Sigma_{1,2}(\sigma_{1,2},\sigma_{1,2}')$ denote the entries $\langle\sigma_{1,2}\vert\Sigma_{1,2}\vert\sigma_{1,2}'\rangle$ of the internal  operators.

On the contrary, (non-normalized) states of two identical particles
in the same spatial region, {\it e.g.} of the form $\mathfrak{S}\big[\vert L,\sigma\rangle\otimes\vert L,\sigma'\rangle\big]$ which, in first quantization, are proportional to
\begin{equation}
\label{ex}
\vert L,\sigma\rangle\otimes\vert L,\sigma'\rangle\pm\vert L,\sigma'\rangle\otimes\vert L,\sigma\rangle\ ,
\end{equation}
correspond, in second quantization, to states proportional to 
$\mathfrak{a}_{L,\sigma}^\dag\mathfrak{a}_{L,\sigma'}^\dag|\textnormal{vac}\rangle$.
However, these states cannot be effectively distinguished by spatial localization since the single-particle states they consist of
are not spatially orthogonal.

The net outcome of effective distinguishability is that
the standard description of distinguishable particles emerges from the general theory of identical ones
through an effective ``freezing'' of suitable degrees of freedom corresponding to single-particle orthogonal states. In this respect, the states considered in this section are paradigmatic: the state in~\eqref{eff.dist.ent.isom} is the prototypical entangled state of two distinguishable particles (a Bell pair), while the state in~\eqref{eff.dist.sep.isom} is separable. 
They can be used as benchmarks for checking whether, once effective distinguishability is implemented, any given definition of identical particle entanglement reduces or not to the standard definition holding for distinguishable ones.

The discussion presented in this Section for two particles can be generalized to many particles and arbitrary Hilbert space dimension, as sketched in~\ref{gen.isom}.

\section{Particle-partitioning}
\label{part-ent}

Entanglement is the strongest form of 
correlations among different degrees of freedom of quantum systems. 
In the case of distinguishable particles,
the degrees of freedom are most naturally those that identify  the particles themselves. We shall refer to such an identification of the relevant degrees of freedom as \textit{particle-partitioning}
which is mathematically expressed by writing the total Hilbert space of the compound system as the tensor product of single-particle Hilbert spaces. 
In the case of two distinguishable particles, the selected degrees of freedom are those of particle $1$, respectively $2$. They are thus associated with the sets of  single-particle operators
\begin{equation}
\label{auxxx1}
\mathcal{A}_1=\{O_1\otimes\mathbbm{1}\}\ ,\quad\hbox{respectively}\quad \mathcal{A}_2=\{\mathbbm{1}\otimes O_2\}\ ,
\end{equation}
With respect to the particle partitioning of such a bipartite system, particle-local observables are products of single-particle operators,
\begin{equation}
\label{loc.product}
O_1\otimes O_2=\big(O_1\otimes\mathbbm{1}\big)\big(\mathbbm{1}\otimes O_2\big)\ .
\end{equation}

This structure is however only allowed when each particle is individually addressable 
so that one may always identify particles by resorting to measurement processes of generic particle-local operators. These operators pertain to specific particles whence they commute with those of other particles and their respective measurements do not interfere with each other.
Particle-partitioning brings  with itself the notion of particle-locality
mentioned in the Introduction whereby local observables are identified with tensor products of single-particle observables.

Since particle partitioning is not permitted when particles are identical, particle-locality might turn out to be physically untenable as locality criterion and a different kind of locality based on more general degrees of freedom than particle ones might then become necessary. Such an extended locality  would in turn entail the extension of entanglement and separability beyond 
particle-entanglement and particle-separability.
\cite{Clifton2001,Zanardi2001,
Barnum2003,Zanardi2004,Barnum2004,Barnum2005,Verch2005,
Harshman2011,Thirring2011,Yngvason2015}.

Since one cannot partition identical particle systems in terms of single (or groups of) particles, what one could do is to choose and partition
degrees of freedom which are not related to any specific particle and accordingly consider
observables associated with the chosen degrees of freedom.
In doing so, one focuses not on sets of observables pertaining to specific particles,
rather on
sets of operators related to selected degrees of freedom, like particle numbers or collective excitations. We shall refer to the selection of generic subsets 
of degrees of freedom as algebraic partitioning. The qualification algebraic refers to the fact that, typically,  the operators associated to different degrees of freedom constitute sets that are closed under summation and multiplication of operators \cite{BratteliRobinson,Strocchi}. 
Algebras and subalgebras are physically preferable to sets and subsets for, given a set of observables, one can always construct other observables by adding and multiplying them.

Notice that, for distinguishable particles, the \textit{algebraic partitioning} is tacitly  identified with particle-partitioning.
As already discussed in the Introduction, bipartite state vectors of distinguishable particles are particle-separable when they give rise to factorizing mean values of particle-local observables.
More in general, two-particle mixed states, represented by density matrices $\rho_{12}$, are particle-separable if they can be expressed as convex combinations, 
namely mixtures, of one-dimensional projections onto particle-separable state vectors.

\begin{definition}[Distinguishable particle entanglement~\cite{Werner1989}]
\label{part-sep}
A bipartite density matrix $\rho_{12}$ of two distinguishable particles described by the
algebraic bipartition $(\mathcal{A}_1,\mathcal{A}_2)$ in \eqref{auxxx1} is separable if and only if, for all $O_{1,2}$,
\begin{equation}
\label{sep}
\textnormal{Tr}\big(\rho_{12}\, O_1\otimes O_2\big)=\sum_j p_j\textnormal{Tr}\Big(\rho_j^{(1)}O_1\Big)\textnormal{Tr}\Big(\rho_j^{(2)}O_2\Big)\ ,
\end{equation}
where $p_j\geqslant 0$ are weights such that $\sum_j p_j=1$,
and $\rho_j^{(1,2)}$ are admissible single-particle states.
All states that do not factorize as above are called entangled.~\footnote{The correlations embodied by~\eqref{sep} are surely particle-local, but not entirely classical. Indeed, beside the statistical correlations contained in the distribution of  weights $p_j$, there can still be quantum correlations, called discord, due to the contributing states $\rho^{(1,2)}_j$ being in general not orthogonal.}
\end{definition}

Since no observables can address individual identical particles, the definition of entanglement
by means of particle correlations is particularly challenging. The main obstacle being that, for identical particles, particle-local
observables as $O_1\otimes O_2$ are only physically acceptable if $O_1= O_2$, so that $O_1\otimes O_2$ becomes trivially symmetric. 
Otherwise, one has to deal with symmetrized observables as in~\eqref{loc.op} which are no longer particle-local.

In trying to cope with this difficulty, different definitions of identical particle-entanglement
have been proposed
that are based on different points of view about what 
entanglement is and what it can be used for. As we shall see in detail, each of these approaches entail a specific notion of particle-locality, even if not explicitly declared. Instead of focusing
on the motivations behind all of them,
we shall analyse their physical consequences.
More specifically, we shall discuss which states are separable and entangled in those
formulations focusing
on the corresponding underlying  notions of particle-locality and on their physical tenability.

For sake of simplicity, we shall restrict to the case of two indistinguishable particles and eliminate the two-particle subscript label ``{12}'' used so far for either pure and mixed states.

As regards locality, the first observation is that, in the case of particle-entanglement, it cannot but concern products of commuting single-particle operators which, unlike for distinguishable particles, cannot be as in~\eqref{auxxx1}. Rather, they must be symmetrized as the observables $\mathcal{P}(O,\mathbbm{1})$ in~\eqref{idensingop}.
Notice that these observables cannot be particle-local in the case of distinguishable particles.
Nevertheless they are the only sensible single-particle observables for identical particles, and one can always focus on the degrees of freedom identified by distinct subsets 
(subalgebras) of them
$\mathcal{A}_j=\{\mathcal{P}(O_j,\mathbbm{1})\}_{O_j\in\mathcal{I}_j}$, where $\mathcal{I}_j$ are sets of selected relevant single-particle observables.

Notice that the condition $[O_1,O_2]=0$ is equivalent to the single-particle operator commutativity; indeed:
\begin{equation}
\label{auxxx2}
\big[\mathcal{P}(O_1,\mathbbm{1}),\mathcal{P}(O_2,\mathbbm{1})\big]=\big[O_1,O_2\big]\otimes\mathbbm{1}+\mathbbm{1}\otimes\big[O_1,O_2\big]\ .
\end{equation}

The sets $\mathcal{I}_j$
cannot contain all single-particle operators $O_j$,
otherwise the two subalgebras $\mathcal{A}_1$ and $\mathcal{A}_2$ obtained by summing and multiplying all operators $\mathcal{P}(O_{1,2},\mathbbm{1})$ 
would both coincide with $M_2(\mathbb{C})\otimes M_2(\mathbb{C})$, where 
$M_n(\mathbb{C})$ is the algebra of complex
$n\times n$ matrices. Typically, one chooses $\mathcal{A}_1$ and $\mathcal{A}_2$ such that
sums and products of their operators generate the whole algebra of which they are commuting subalgebras, the standard example being, for two qubits,
$\mathcal{A}_1=M_2(\mathbbm{C})\otimes\mathbbm{1}$,  $\mathcal{A}_2=\mathbbm{1}\otimes M_2(\mathbbm{C})$ that generate $M_4(\mathbb{C})$.

Once an algebraic bipartition based on two subsets of selected single-particle degrees of freedom has been chosen, the resulting locality criterion can be expressed as follows:

\begin{definition}
\label{def-idloc}
Identical particle
operators are particle-local with respect to
a given algebraic particle-bipartition $(\mathcal{A}_1,\mathcal{A}_2)$
if they have the form
\begin{equation}
\label{POOP}
A_1 A_2 \quad \textnormal{with} \quad A_j\in\mathcal{A}_j=\{\mathcal{P}(O_j,\mathbbm{1})\}_{O_j\in\mathcal{I}_j},
\end{equation}
with $j=1,2$ and $[O_1,O_2]=0$.
\end{definition}

According to the previous definition and as discussed in the Introduction, 
 two-particle state vectors  $\vert\psi\rangle$ are particle-separable with respect to the chosen 
partition $(\mathcal{A}_1,\mathcal{A}_2)$ of single particle observables if they fulfil
the factorization condition
\begin{equation} 
\label{auxxx30}
\langle\psi|A_1A_2|\psi\rangle=\langle\psi|A_1|\psi\rangle\langle\psi|A_2|\psi\rangle\ ,
\end{equation}
for all operators $A_j\in\mathcal{A}_j$, $j=1,2$. 
For mixed states, such an equality
generalizes to (compare with \eqref{sep})
\begin{equation}
\label{sep2}
\textnormal{Tr}\big(\rho\, A_1\, A_2\big)=\sum_j p_j\,\textnormal{Tr}\Big(\rho_j^{(1)}A_1\Big)\textnormal{Tr}\Big(\rho_j^{(2)}A_2\Big)\ ,
\end{equation}
with $A_1\in\mathcal{A}_1$, $A_2\in\mathcal{A}_2$, $p_j\geqslant 0$ such that $\sum_j p_j=1$,
and $\rho_j^{(1)}$, $\rho_j^{(2)}$ admissible states.

We shall now consider in detail various possible definitions of indistinguishable particle-entanglement,
reflecting different points of view about what non-local quantum correlations among particles should amount to.
By focusing on what entangled states are declared to be in the various approaches present in the literature, one identifies four general classes, that 
will be listed under the caption \textit{entanglement-I, -II, -III and -IV};
accordingly we shall speak of \textit{separability-I, -II, -III and -IV} and of states that are \textit{entangled-I, -II, -III and -IV}, respectively \textit{separable-I, -II, -III 
and -IV}. In all these definitions, no notion of locality  of observables has been explicitly discussed. 
However, since the four formulations make use of a first quantization formalism,
the reference to particles, rather than to generic degrees of freedom, as constituent parties of compound systems is implicitly assumed. The locality criterion that emerges from these choices is then precisely the one 
in Definition~\ref{def-idloc}.
Our aim is to check to what extent these entanglement notions
are compatible with the three physical criteria introduced in Section~\ref{Intro}.

\subsection{Entanglement-I}\label{fact.sect}

The first approach to identical particle-entanglement is based on the tensor product factorization of single-particle states, that is on particle-locality as for distinguishable particles.

\begin{definition}[Entanglement-I]
\label{fact.ent}
Pure Bosonic states of $N$-identical particle systems are separable if all particles occupy the same single-particle state:
$|\psi\rangle^{\otimes N}$  for any single-particle state $|\psi\rangle$. All other states are entangled.
\end{definition}

\noindent
Several entanglement criteria, entanglement measures and entanglement witnesses \cite{Paskauskas2001,Eckert2002,Wang2005,Zhou2008,Kus2009,
Kotowski2010,Grabowski2011,Grabowski2011-2,Sawicki2012,
Hyllus2012,Oszmaniec2013,Wasak2014,Killoran2014,Oszmaniec2014,
Sawicki2014,Reusch2015,Rigolin2016,GarciaCalderon2017,
Morris2019}, together with protocols for entanglement manipulation \cite{Karczewski2019}, and connections with condensed matter systems \cite{Naudts2007,Zozulya2008,Haque2009,Kunkel2018} are presented and discussed in connection with the previous definition.

Definition \ref{fact.ent} has been formulated for Bosons; however, it can be  extended to Fermions whereby it implies that all Fermionic states are entangled-I, due to their anti-symmetrization \cite{Cavalcanti2007,Ichikawa2008}.

\subsubsection{Local operators} \label{loc.op.fact}

As already observed, being
entanglement-I formulated in reference to  particles, the underlying locality criterion is that of Definition~\ref{def-idloc}, namely that local observables are products of commuting single-particle operators. It follows that
entanglement-I conflicts with the factorization property in~\eqref{auxxx30}; indeed, one can exhibit expectations with respect to separable-I states that fail to factorize on products of observables coming from commuting subsets of single-particle operators. More specifically, in the following it is shown that, given all possible
products of operators $A_i=\mathcal{P}(O_i,\mathbbm{1})$, $i=1,2$, belonging to commuting subsets, 
there always exists a separable-I pure state giving rise to expectations that do not factorize on some of them.

Consider two arbitrary single-particle operators $\mathcal{P}(O_i,\mathbbm{1})$, $i=1,2$, such that $[O_1,\,\,O_2]=0$,  and their product
\begin{align}
\mathcal{P}(O_1,\mathbbm{1})\mathcal{P}(O_2,\mathbbm{1})= & \, O_1 O_2\otimes\mathbbm{1}+\mathbbm{1}\otimes O_1 O_2 \nonumber \\
& +O_1\otimes O_2+O_2\otimes O_1.
\label{extended.loc}
\end{align}
Commutativity of $O_{1,2}$ implies that  they  have a common eigenbasis $\{|e_\lambda\rangle\}_\lambda$ corresponding to eigenvalues $\{o^{(1)}_\lambda\}_\lambda$ for $O_1$, and $\{o^{(2)}_\lambda\}_\lambda$ for $O_2$. Let us now consider
separable-I states of the form $|\psi\rangle\otimes|\psi\rangle$ with
\begin{equation}
|\psi\rangle=\frac{|e_\lambda\rangle+|e_\kappa\rangle}{\sqrt{2}},
\end{equation}
and $\lambda\neq\kappa$. Then, one computes
\begin{align}
& \langle\psi\otimes\psi|\mathcal{P}(O_1,\mathbbm{1})\mathcal{P}(O_2,\mathbbm{1})|\psi\otimes\psi\rangle- \nonumber \\
& \langle\psi\otimes\psi|\mathcal{P}(O_1,\mathbbm{1})|\psi\otimes\psi\rangle\langle\psi\otimes\psi|\mathcal{P}(O_2,\mathbbm{1})|\psi\otimes\psi\rangle \nonumber \\
& =\frac{1}{2}\big(o^{(1)}_\lambda-o^{(1)}_\kappa\big)\big(o^{(2)}_\lambda-o^{(2)}_\kappa\big)\ .
\label{fact.id.fact.ent}
\end{align}
Therefore, the above correlation functions factorize into products of single-particle expectations if and only if~\eqref{fact.id.fact.ent} vanishes, namely if and only if either $o^{(1)}_\lambda=o^{(1)}_\kappa$ or $o^{(2)}_\lambda=o^{(2)}_\kappa$. Since the indices $\lambda$ and $\kappa$ are arbitrary, factorization holds if and only if either $O_1$ or $O_2$ is proportional to the identity.
In conclusion, for whatever choice of non-trivial commuting subsets of observables $\mathcal{A}_i=\{\mathcal{P}(O_i,\mathbbm{1})\}_{O_i\in\mathcal{I}_i}$ with $i=1,2$, there are always
separable-I states that do not fulfil the factorization of local expectations in \eqref{auxxx30}.

Additional remarks regarding entanglement-I are in order.
The incompatibility between entanglement-I and particle-locality is also highlighted by
the fact that single-particle operators $\mathcal{P}(O,\mathbbm{1})$ can map separable-I states into
entangled-I states.
This fact is taken by some as a witness of the non-locality of the operators $\mathcal{P}(O,\mathbbm{1})$. This argument is however inconsistent since operators of that form are
the only physically relevant single-particle operators and, as stated
in the Introduction, entanglement must be derived from a chosen locality criterion concerning products of single-particle operators and not vice versa.

Entanglement-I can also be seen as the restriction of distinguishable particle-entanglement in Definition \ref{part-sep} to the symmetric subspace. One is therefore tempted to state that the relevant particle-locality is induced by the commuting subalgebras $\mathcal{A}_1=\{O_1\otimes\mathbbm{1}\}$ and $\mathcal{A}_2=\{\mathbbm{1}\otimes O_2\}$.
We stress again that $(\mathcal{A}_1,\mathcal{A}_2)$ cannot be a bona fide algebraic bipartition. Indeed, operators in either $\mathcal{A}_1$ or $\mathcal{A}_2$ are not
compatible with particle indistinguishability (as well as many of the formally local operators $A_1 A_2$ with $A_j\in\mathcal{A}_j$, $j=1,2$).
In this setting, the only
operators compatible with entanglement-I would be ``collective'' ones, i.e. $O\otimes O$ which are however truly local only in the distinguishable particle setting; indeed, they cannot be written as product of symmetrized single-particle operators of the form $\mathcal{P}(O_{1,2},\mathbbm{1})$ with $[O_1\,,\,O_2]=0$.

\subsubsection{Effective distinguishability} \label{eff.dist.fact}

In the following, as a typical context for effective distinguishability, we consider again particles described by an ``external'', or spatial, degree of freedom with values $L,R$, corresponding to being confined within non-overlapping left and right volumes, and an ``internal'', spin, degree of freedom $\sigma$, with values $\uparrow,\downarrow$ (see Section~\ref{eff.dist}).
One can then easily construct two-particle states not of the form $\vert\psi\rangle\otimes\vert\psi\rangle$, therefore
entangled-I, that, by using spatial orthogonality, can be made effectively correspond to distinguishable particle separable states ({\it e.g.} see ~\eqref{eff.dist.sep} and ~\eqref{eff.dist.sep.isom} in Section \ref{eff.dist}).
Therefore, distinguishable particle-entanglement cannot be recovered from entanglement-I.
In addition, mean values of generic operators that are local in the framework of effectively distinguishable particles, i.e. those
of the form $\mathcal{P}(P_L\otimes\Sigma_1,P_R\otimes\Sigma_2)$
as discussed in Section \ref{eff.dist}, do not fulfil the factorization \eqref{auxxx30} because of the argument developed just above.

\begin{figure}[htbp]
\centering
\includegraphics[width=\textwidth]{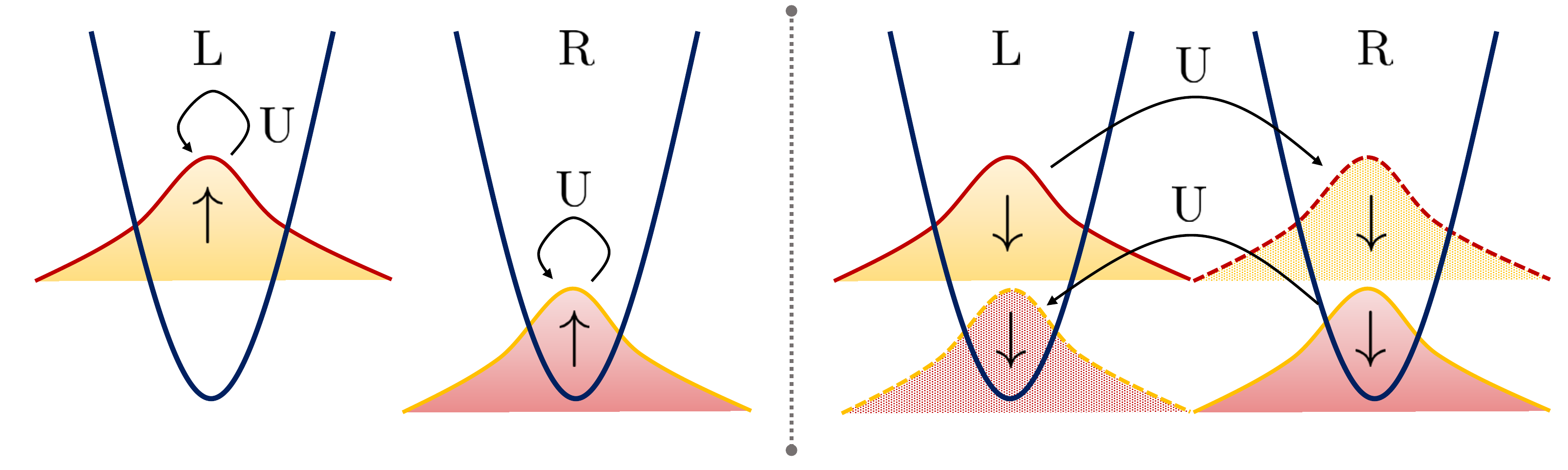}
\includegraphics[width=\textwidth]{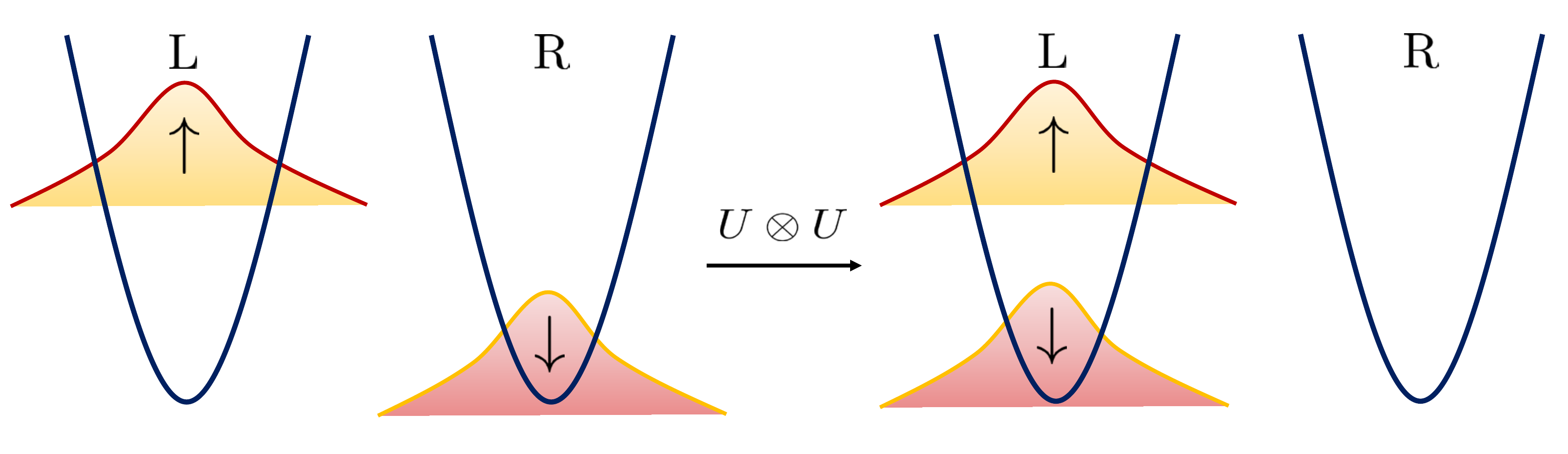}
\caption{Pictorial action of the operators $U$ (upper panel) and $U\otimes U$ (lower panel).}
\label{fig.U}
\end{figure}

Such an inconsistency is emphasised in the physical situation where identical particles that have been effectively distinguished, by confining them in disjoint spatial regions, are moved close together. The paradox here is that entanglement-I is created by a procedure which is described by an operator of the form $U\otimes U$, which cannot create entanglement-I for it preserves the form $\vert\psi\rangle\otimes\vert\psi\rangle$ of separable-I state vectors.
In particular, consider the single-particle operator (see Figure \ref{fig.U})
\begin{eqnarray}
\nonumber
U&=&|L,\uparrow\rangle\langle L,\uparrow|+|R,\uparrow\rangle\langle R,\uparrow| \\
&&\hskip 1cm
+|R,\downarrow\rangle\langle L,\downarrow| +
|L,\downarrow\rangle\langle R,\downarrow|\ . \label{UU}
\end{eqnarray}
Then, acting with $U\otimes U$ on states of the form~\eqref{eff.dist.sep} yields
\begin{equation}
U\otimes U \, |\phi^\textnormal{id}_1\rangle=
\mathfrak{S} \big[ |L,\uparrow\rangle\otimes|L,\downarrow\rangle \big] \ ,
\label{v1v2id}
\end{equation}
which is an entangled-I state.

Since $U\otimes U$ cannot generate entanglement-I,
the entanglement in the state \eqref{v1v2id} must
be already present in the initial state $|\phi^\textnormal{id}_1\rangle$. This possibility is however incompatible with distinguishable particle-entanglement,
because $|\phi^\textnormal{id}_1\rangle$ becomes the separable state $|\phi^\textnormal{dist}_1\rangle$ in~\eqref{eff.dist.sep.isom} once the particle have been distinguished by confining them in the $L,R$ spatial regions.
	
In conclusion, as regards the compatibility of entanglement-I with effective distinguishability, it is important to notice that entangled states of effectively distinguishable particles can be experimentally prepared starting from Bosonic permanents or Slater determinants, which are entangled-I states.
These procedures rely on inter-particle interactions either to move particles away from each other or to manipulate their states when the latter
are spatially localized within a same region \cite{Cavalcanti2007,Hayes2007,Babcock2008,Kaufman2015,
Deacon2015,Tan2015,Lange2018,Blasiak2019,Karczewski2019}. Entanglement thus seems to be more the result of particle interactions
 than of
the (anti-)symmetrization of the initial state as implied by
\hbox{separability-I}.
If producing and extracting entanglement solely from \hbox{(anti-)symmetrization} were possible, then identical particles would provide a source of entanglement with infinite capacity.

\subsubsection{Information resources} \label{info.res.fact}

All evidence gathered so far in the literature shows no conflict of
entanglement-I with this criterion. However, 
operational applications of
entanglement-I are rather limited, being restricted to interferometric phase estimation protocols~\cite{Pezze2009}.

\subsection{Entanglement-II} \label{perm.sect}

A rather different point of view with respect to
entanglement-I can be developed by assuming that
the entanglement generated by Bosonic symmetrization and Fermionic antisymmetrization
is in fact a mere mathematical artefact of no use in information processing. Indeed, the impossibility of addressing individual particles puts a limit on the usability of some states, entangled according to
entanglement-I, for teleportation \cite{Marinatto2001,
Molotkov2010,
Marzolino2015,Marzolino2016}, rendering them somewhat inert.
For these reasons, the following definition
is introduced originally for Fermions, and then extended also to Bosons.

\begin{definition}[Entanglement-II]
\label{perm.ent}
Pure separable states of Bosons (Fermions) are the symmetrization (anti-symmetrization) of tensor products 
$\bigotimes_{j=1}^N|\psi_j\rangle$ of single-particle states $|\psi_j\rangle$, that are pairwise either orthogonal, 
$\langle\psi_j|\psi_{j'}\rangle=\delta_{j,j'}$, or equal, $\langle\psi_j|\psi_{j'}\rangle=1$.
All other states are entangled.
\end{definition}

This definition underlies different approaches \cite{Herbut2001,Schliemann2001,Paskauskas2001,Eckert2002,
Li2001,Ghirardi2002,Ghirardi2003,Fang2003,Ghirardi2004,
Ghirardi2004-2,Ghirardi2005,Wang2005,Levay2008,Vrana2009,Kus2009,Plastino2009,Zander2010,
Kotowski2010,Grabowski2011,Grabowski2011-2,Sawicki2012,Zander2012,
Oszmaniec2013,Iemini2013,Iemini2013-2,Ladyman2013,Iemini2014,
Oszmaniec2014,Tichy2014,Sawicki2014,Reusch2015,Lourenco2019,
Gigena2020} that use entanglement criteria and entanglement measures and apply them to entanglement manipulation \cite{Bouvrie2017} and to several condensed matter systems \cite{Ramsak2006,Buscemi2006,Zhu2006,Buscemi2007,Buscemi2007-2,
Buscemi2007-3,Naudts2007,Buscemi2008,Yanez2010,Esquivel2011,
Majtey2012,Bouvrie2012,Bouvrie2014,Lin2015,Iemini2015}.

In compound multi-partite states as in Definition \ref{perm.ent}, each particle is in a single-particle pure state, namely it possesses a complete set of properties
(eigenvalues of a complete set of single-particle commuting operators),
but in a permutationally invariant way, so that we cannot know which particle is in which state \cite{Ghirardi2002}. 
Thus, compared to entanglement-I, entanglement-II significantly extends the set of separable states.

Notice that, according to the above Definition, \hbox{(anti-)symmetrized} states 
with single-particle contributions that are neither mutually orthogonal, nor equal, are  entangled-II.

\subsubsection{Local operators} \label{loc.op.perm}

As for entanglement-I, the implicit locality criterion underlying entanglement-II refers to particles
and not to generic degrees of freedom. It follows that the underlying notion  of local observables is the one given in Definition~\ref{def-idloc}.
Then, the argument proposed in Section \ref{loc.op.fact} represents an inconsistency also for entanglement-II because states that are separable according to entanglement-I are  also separable according to entanglement-II.
In the following, this argument is adapted to states that are entangled-I but separable-II.

Consider two arbitrary commuting single-particle operators, $\mathcal{P}(O_1,\mathbbm{1})$ and $\mathcal{P}(O_2,\mathbbm{1})$; using the same notation introduced in Section \ref{loc.op.fact}, let $\{|e_\lambda\rangle\}_\lambda$ be common eigenvectors of $\mathcal{P}(O_{1,2},\mathbbm{1})$ corresponding to eigenvalues $\{o_\lambda^{(1)}\}_\lambda$ and $\{o_\lambda^{(2)}\}_\lambda$ of $O_1$ and $O_2$ respectively. Consider the
separable-II  states
\begin{equation} \label{counterex.loc.perm}
|\phi\rangle=\sqrt{2} \, \mathfrak{S}\big[|\psi_1\rangle\otimes|\psi_2\rangle\big]=\frac{|\psi_1\rangle\otimes|\psi_2\rangle\pm|\psi_2\rangle\otimes|\psi_1\rangle}{\sqrt{2}},
\end{equation}
with
\begin{equation} \label{counterex12}
|\psi_1\rangle=\frac{|e_\lambda\rangle+|e_\kappa\rangle}{\sqrt{2}}, \qquad |\psi_2\rangle=|e_\mu\rangle,
\end{equation}
and $\lambda$, $\kappa$ and $\mu$ all different. Then
\begin{align}
& \langle\phi|\mathcal{P}(O_1,\mathbbm{1})\mathcal{P}(O_2,\mathbbm{1})|\phi\rangle-\langle\phi|\mathcal{P}(O_1,\mathbbm{1})|\phi\rangle\langle\phi|\mathcal{P}(O_2,\mathbbm{1})|\phi\rangle \nonumber \\
& =\frac{1}{4}\big(o^{(1)}_\lambda-o^{(1)}_\kappa\big)\big(o^{(2)}_\lambda-o^{(2)}_\kappa\big)\ .
\label{fact.id.perm.ent}
\end{align}

By the same argument developed in Section \ref{loc.op.fact}, expectations of the form \hbox{$\langle\phi|\mathcal{P}(O_1,\mathbbm{1})\mathcal{P}(O_2,\mathbbm{1})|\phi\rangle$} involving products of single-particle operators,
factorize into products of single 
particle expectations if and only if either $o^{(1)}_\lambda=o^{(1)}_\kappa$, or $o^{(2)}_\lambda=o^{(2)}_\kappa$ so that,
by the arbitrariness of $\lambda$ and $\kappa$, if and only if either $O_1$ or $O_2$ is proportional to the identity matrix. Therefore, entanglement-II is not compatible with
identical particle locality.

As an additional remark, let us consider the state \eqref{counterex.loc.perm} with $|\psi_1\rangle$ and $|\psi_2\rangle$ that are both
orthogonal superpositions of 
two eigenstates $|e_\lambda\rangle$ and $|e_\kappa\rangle$, e.g. $|\psi_1\rangle$ as in \eqref{counterex12} and $|\psi_2\rangle=\big(|e_\lambda\rangle-|e_\kappa\rangle\big)/\sqrt{2}$. In this case, the quantity at the left-hand-side of~\eqref{fact.id.perm.ent} is proportional to the right-hand-side only for Bosons, while it vanishes for Fermions, whence the necessity of the use of a third common eigenstate $|e_\mu\rangle$ of $O_{1,2}$; indeed, for a two-dimensional single-particle Hilbert space there is only one Fermionic two-particle state which is separable-II.

\subsubsection{Effective distinguishability} \label{eff.dist.perm}

At first glance,
entanglement-II appears compatible with effective distinguishability because, according to the isomorphisms discussed in Section \ref{eff.dist},
separable states of distinguishable particles correspond to permanents for Bosons or Slater determinants for Fermions and,
therefore to separable-II states.
Indeed, in \cite{Ghirardi2002} the factorization \eqref{auxxx30} was checked to hold
for states with precisely one particle localized within the left region and the other one within the right region, as discussed in Section~\ref{eff.dist} (see~(\ref{eff.dist.sep}) and~\ref{eff.dist.ent})). Furthermore, the argument developed there involves operators of the form $\mathcal{P}(O_{1,2},\mathbbm{1})$ with single particle operators $O_1=P_L\otimes\Sigma$ and $O_2=P_R\otimes\Sigma'$ such that $[\mathcal{P}(O_1,\mathbbm{1}),\mathcal{P}(O_2,\mathbbm{1})]=0$.
With this choice, $\mathcal{P}(O_1,O_2)$ are local operators in the framework of effective distinguishability discussed in Section~\ref{eff.dist}
and also particle-local operators according to
Definition~\ref{def-idloc}, for
\begin{equation}
\mathcal{P}(O_1,O_2)=\mathcal{P}(O_1,\mathbbm{1})\mathcal{P}(O_2,\mathbbm{1})\ .
\label{Ghi2}
\end{equation}

\begin{figure}[htbp]
\centering
\includegraphics[width=0.7\textwidth]{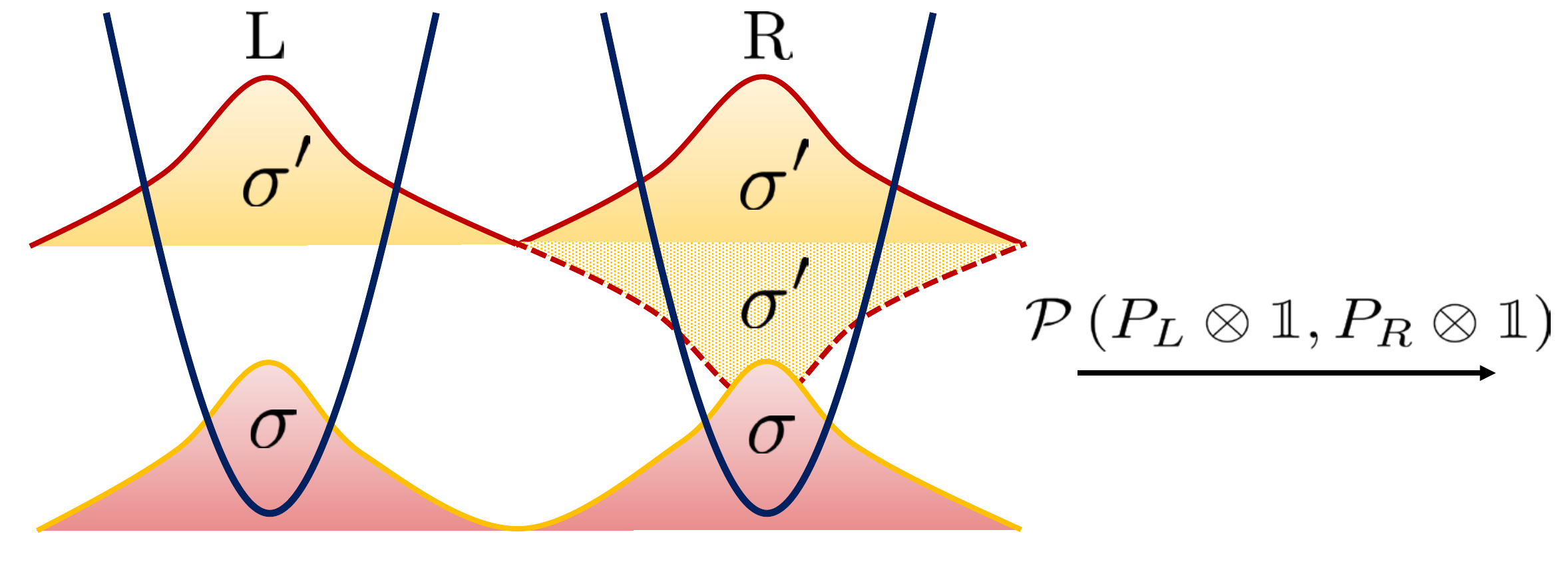}
\includegraphics[width=\textwidth]{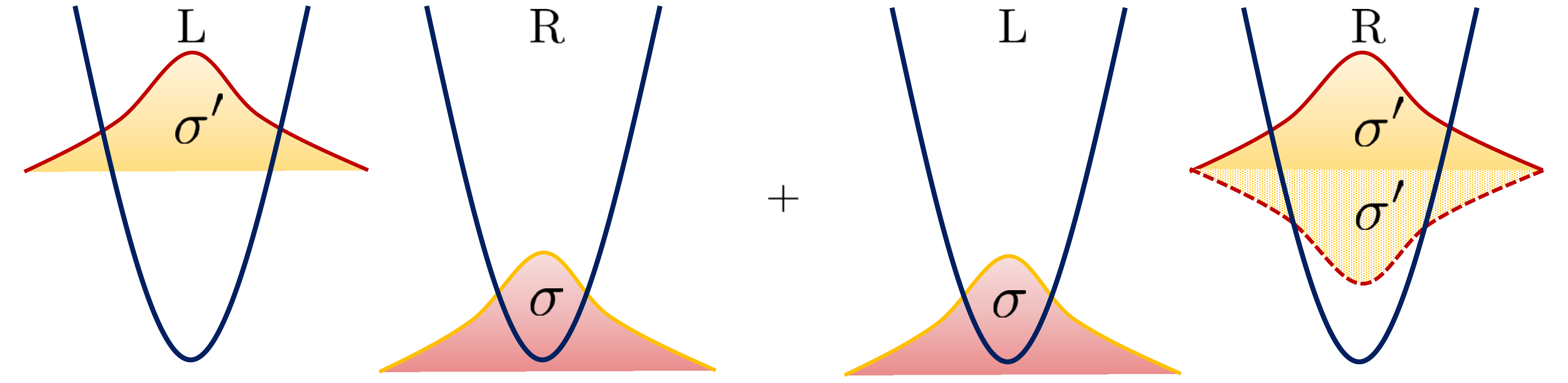}
\caption{Pictorial action of the projection $\mathcal{P}(P_L\otimes\mathbbm{1},P_R\otimes\mathbbm{1})$.}
\label{fig.proj}
\end{figure}

Nevertheless, the theory of entanglement-II developed in~\cite{Ghirardi2002} is not restricted to effective distinguishable particles,
and lifting this restriction exposes another incompatibility
of entanglement-II.
Such an incompatibility can be understood by expanding the approach in \cite{Tichy2013} whereby one distinguishes an \textit{a priori} entanglement, that is equivalent to entanglement-II, from a so-called \textit{detector-level} entanglement. The latter is the entanglement that emerges after a measurement has implemented effective distinguishability by ``freezing'' suitable degrees of freedom, e.g. through the operator $\mathcal{P}\big(P_L\otimes\mathbbm{1},P_R\otimes\mathbbm{1}\big)$ (see Figure \ref{fig.proj}) or by limiting observables to linear combinations of $\mathcal{P}\big(P_L\otimes\Sigma,P_R\otimes\Sigma'\big)$. As we elaborate below, such a \textit{freezing by measuring} maps a priori separable-II states into detector-level entangled states despite being particle-local in the framework of effective distinguishability. Moreover, the generated entanglement depends on internal degrees of freedom although these latter are left unaffected by the operations leading to the ``freezing'' of degrees of freedom as they are decoupled from the spatial degrees of freedom.

In order to be more specific, consider the state
\begin{align}
|\zeta_+\rangle = & \sqrt{\frac{2}{1+|\langle\sigma|\sigma'\rangle|^2}} \, \mathfrak{S}\bigg[\Big(\vert +\rangle\otimes\vert\sigma\rangle\Big)
\otimes\Big(\vert+\rangle\otimes\vert\sigma'\rangle\Big)\bigg] \nonumber \\
= & \Big(\vert+\rangle\otimes\vert+\rangle\Big)\otimes|\xi_\textnormal{internal}\rangle \, ,
\label{sigmagamma1}
\end{align}
obtained by separating
the tensor product
into spatial, $\displaystyle\vert+\rangle=\frac{\vert L\rangle+\vert R\rangle}{\sqrt{2}}$, and internal contributions
\begin{equation}
\vert\xi_\textnormal{internal}\rangle= \sqrt{\frac{2}{1+|\langle\sigma|\sigma'\rangle|^2}} \,
\mathfrak{S} \big[ \vert\sigma\rangle\otimes\vert\sigma'\rangle\big] \, .
\end{equation}
Since the spatial states are not orthogonal, the state in equation \eqref{sigmagamma1}
describes particles that
cannot be effectively distinguished by their  spatial degrees of freedom. Moreover, according to Definition~\ref{perm.ent}, the state $|\zeta_+\rangle$ is
separable-II if the two single-particle states are the same, $|\langle \sigma | \sigma' \rangle| = 1$, or if they are orthogonal, $\langle \sigma | \sigma' \rangle = 0$.

On the other hand, the particle-local observable, as in Definition \ref{def-idloc},
$$
\mathcal{P}(P_L\otimes\mathbbm{1},P_R\otimes\mathbbm{1})=\mathcal{P}(P_L,1)\mathcal{P}(P_R,1)=P_L\otimes P_R+P_R\otimes P_L
$$ 
turns
$\vert\zeta_+\rangle$ into a state proportional to (see Figure \ref{fig.proj})
\begin{equation}
\mathfrak{S}\Big[\Big(|L\rangle\otimes|\sigma\rangle\Big)\otimes\Big(|R\rangle\otimes|\sigma'\rangle\Big)
+\Big(|L\rangle\otimes|\sigma'\rangle\Big)\otimes	\Big(|R\rangle\otimes|\sigma\rangle\Big)\Big] \, ,
\end{equation}
which is
separable-II only if $|\langle\sigma\vert\sigma'\rangle|=1$, entangled-II otherwise.

The same phenomenon occurs considering 
\begin{equation}
\label{sigmagamma2}
|\zeta_-\rangle =\frac{1}{\sqrt{2}} \, \mathfrak{S}\Big[\Big(\vert+\rangle\otimes|\sigma\rangle\Big)\otimes\Big(\vert-\rangle\otimes|\sigma'\rangle\Big)\Big]
\end{equation}
where $\displaystyle\vert\pm\rangle=\frac{\vert L\rangle\pm\vert R\rangle}{\sqrt{2}}$.
Unlike the state $\vert\zeta_+\rangle$  in~\eqref{sigmagamma1}, the above state describes particles that are effectively distinguishable by projecting onto the spatially orthogonal, but not spatially localized, states $|\pm\rangle$ (see Figure~\ref{fig.dist}).
Furthermore, the state \eqref{sigmagamma2} is always
separable-II because of $\langle+\vert-\rangle=0$, while
\hbox{$\mathcal{P}(P_L\otimes\mathbbm{1},P_R\otimes\mathbbm{1})$} turns it into a state
proportional to
\begin{equation}
\mathfrak{S}\Big[\Big(\vert L\rangle\otimes\vert\sigma\rangle\Big)\otimes\Big(\vert R\rangle\otimes\vert\sigma'\rangle\Big) -\Big(\vert L\rangle\otimes\vert\sigma'\rangle\Big)\otimes\Big(\vert R\rangle\otimes\vert\sigma\rangle\Big)\Big] \, ,
\end{equation}
which is entangled-II unless $|\langle\sigma\vert\sigma'\rangle|=1$.

Therefore, the action of the operator $\mathcal{P}(P_L\otimes\mathbbm{1},P_R\otimes\mathbbm{1})$, which implements effective distinguishability and is particle-local (see Section \ref{eff.dist}), creates entanglement-II for certain values of the internal degrees of freedom, although the latter are not 
affected by the operation associated with $\mathcal{P}(P_L\otimes 1,P_R\otimes 1)$. Also, as explicit in \eqref{sigmagamma1},
the internal degrees of freedom are independent from the spatial degrees of freedom.
The fact that the entanglement generation relies upon independent degrees of freedom discloses a further inconsistency of entanglement-II.

The use of spatial and internal degrees of freedom enables us to emphasize some physical consequences of this inconsistency. 
Indeed,
acting with the symmetric observable $\mathcal{P}(P_L\otimes\mathbbm{1},P_R\otimes\mathbbm{1})$ corresponds to placing two detectors at positions $L$ and $R$, respectively, and to post-selecting double counting events. Nevertheless, the above operation
performed upon identical particles without internal degrees of freedom is mathematically associated with the action of the observable $\mathcal{P}(P_L,P_R)$. The latter is a one-dimensional projector onto a
separable-II state in the (anti-)symmetrized Hilbert space, and thus never generates entanglement-II. Therefore, post-selecting double counting events from detections at positions $L$ and $R$ would unnaturally provide entanglement solely because additional (internal) degrees of freedom exist.

\subsubsection{Information resources} \label{info.res.perm}

Entanglement-II identifies states that can be used as quantum resources
in
teleportation protocols~\cite{Marinatto2001,Molotkov2010}. 
The notion of
entanglement-II is however not sufficient to explain
quantum enhancement in interferometric phase estimation.

Consider indeed the estimation of the relative phase between two arms of an interferometer. In the case of distinguishable particles, enhanced quantum performances overcoming classical ones require entanglement, whereas, for identical particle quantum advantages also occur without initial entanglement. 
Indeed, consider the $N$ Boson particle state
\begin{equation} \label{Fock-states}
\mathcal{S}  \left[ |0\rangle^{\otimes k}\otimes|1\rangle^{\otimes(N-k)} \right], \qquad 0<k<N,
\end{equation}
where $|0\rangle,|1\rangle$ are the eigenvectors of a spin-like observable, as $\sigma^z$. Suppose now to inject such a state into an interferometer whose action is modelled by the unitary $\bigotimes_j e^{i\theta\sigma_j^x}$ which rotates the spin-like degree of freedom around the $x$ axis.
Notice that the state \eqref{Fock-states} is
separable-II and that the interferometer action cannot generate
entanglement-II.
Indeed,
\begin{align}
& \bigotimes_j e^{i\theta\sigma_j^x}
\mathcal{S}\left[|0\rangle^{\otimes k}\otimes|1\rangle^{\otimes(N-k)}\right]= \nonumber \\
& =\mathcal{S}\left[\Big(e^{i\theta\sigma^x}|0\rangle\Big)^{\otimes k}\otimes\Big(e^{i\theta\sigma^x}|1\rangle\Big)^{\otimes(N-k)}\right]\ .
\end{align}

Despite the absence of particle non-locality, in~\cite{Benatti2010,Benatti2011,Argentieri2011,Benatti2014} it has been shown that there is a quantum advantage in the accuracy in the estimation of the phase $\theta$.
Therefore,
entanglement-II fails to identify the non-local resources in this interferometric set-up.

\subsection{Entanglement-III} \label{dist.sect}

A third approach  to identical particle entanglement identifies quantum correlations of identical particles
by directly relating them to those of distinguishable particles.
Consequently, also this point of view is implicitly  based on a locality criterion that refers to particles as in Definition~\ref{def-idloc}.

This third approach proceeds by starting from the Hilbert space $\mathfrak{S} \left[ \mathcal{H}^{\otimes N} \right]$ of $N$-identical particles. Given a partition of the single-particle Hilbert space $\mathcal{H}$ into two orthogonal subspaces $V_{1,2}$, $\mathcal{H}=V_1\bigoplus V_2$, let $\Pi_{n_j}^{(j)}$, $j=1,2$, project onto the subspace of $\mathfrak{S} \left[ \mathcal{H}^{\otimes N} \right]$ with $n_1$, respectively $n_2$ particles  described by states in  $V_1$, respectively $V_2$, namely  onto $\mathfrak{S}\big[V_1^{\otimes n_1}\otimes V_2^{\otimes n_2}\big]$.
As shown in Section~\ref{eff.dist} and \ref{gen.isom}, this subspace is isomorphic to the Hilbert space $\mathfrak{h}_{n_1}\otimes \mathfrak{h}_{n_2}$ with 
$\mathfrak{h}_{n_j}=\mathfrak{S} \big[ V_j^{\otimes n_j} \big]$.

\begin{definition}[Entanglement-III] \label{loc.ssr.ent}
An $N$-particle pure state $\vert\psi\rangle\in\mathfrak{S} \left[ \mathcal{H}^{\otimes N} \right]$ is entangled-III if there exist positive integers $n_1$ and 
$n_2=N-n_1$ such that, after normalization,
$\Pi_{n_1}^{(1)}\Pi_{n_2}^{(2)}\vert\psi\rangle$ is entangled as a state in $\mathfrak{h}_{n_1}\otimes \mathfrak{h}_{n_2}$.
\end{definition}

The consequences of Definition~\ref{loc.ssr.ent} have been
studied by several authors \cite{Wiseman2003,Vaccaro2003,Fang2003,White2009,Ichikawa2010,
Sasaki2011,Buscemi2011,Benedetti2012,Sasaki2013,Dalton2014,
Dalton2017-1,Dalton2017-2,Wasak2018}, and applied in particular to condensed matter systems
\cite{Dowling2006,Laskowski2015,Beggi2016,Barghathi2018,
Barghathi2019}

The motivation for this formulation of particle-entanglement originates from a rather specific physical setting requiring information transmission from a system of identical particles to a quantum register made of distinguishable qubits, 
usually described by tensor products of single-particle Hilbert spaces with fixed numbers of particles in suitably chosen subgroups.
This condition on one hand confirms that the locality criterion underlying entanglement-III is also related to the particle picture
as in Definition~\ref{def-idloc} and, on the other hand, that it can
be implemented by selecting degrees of freedom and corresponding observables associated with the confinement of individual particle states to orthogonal subspaces, say $V_1$ and $V_2$, of the single-particle Hilbert space. Therefore, these orthogonal sectors identify distinguishable groups of particles, each consisting of $n_j$
of them.

\subsubsection{Local operators} \label{loc.op.dist}

When the condition for entanglement-III is enforced for all possible $n_1$ and $n_2$, the particle-locality notion in agreement with
entanglement-III is entailed by the commuting subalgebras ($j=1,2$)
\begin{equation} \label{subalg-part-I}
\mathcal{A}_j=\big\{\mathcal{P}(O_j,\underbrace{\mathbbm{1},\dots,\mathbbm{1}}_{N-1 \textnormal{ times}}) \, : \, \textnormal{supp}(O_j)\subseteq V_j\big\},
\end{equation}
such that
$\displaystyle\sum_{n_j=0}^\infty\,\Pi_{n_j}^{(j)}\mathcal{A}_j\,\Pi_{n_j}^{(j)}=\mathcal{A}_j$.
These subalgebras are generated by the operators $\mathcal{P}(O_j,\mathbbm{1},\dots,\mathbbm{1})$ that generalize the single-particle operators in~\eqref{idensingop} to more than two identical particles and $\textnormal{supp}(O_j)$ is the support of the operator $O_j$, namely, $O_j$ vanishes on the subspace orthogonal to $\textnormal{supp}(O_j)$.
When
the definition of \hbox{entanglement-III} only refers to specific values of $n_1$ and $n_2$, the relevant commuting subalgebras are $\mathcal{A}^{(j)}_{n_j}=\Pi_{n_j}^{(j)}\mathcal{A}_j\Pi_{n_j}^{(j)}$ with $j=1,2$, and $\Pi_{n_j}^{(j)}$ the projector onto the subspace with $n_j$ particles, supported by the single-particle subspace $V_j$.

In the following, we prove the agreement of entanglement-III with the particle-locality notion inherited from the choice of the subalgebras \eqref{subalg-part-I}. The case of subalgebras $\mathcal{A}^{(j)}_{n_j}$ is then straightforward since $\mathcal{A}^{(j)}_{n_j}\subset\mathcal{A}_{j}$.

For a vector state $|\psi\rangle$, separability-III can be rephrased in terms of states
\begin{equation} \label{sep-sub-C}
|\psi_{n_1,n_2}\rangle=\frac{\Pi_{n_2}^{(2)}\Pi_{n_1}^{(1)}|\psi\rangle}{\sqrt{p_{n_1,n_2}}}, \qquad p_{n_1,n_2}=\langle\psi|\Pi_{n_2}^{(2)}\Pi_{n_1}^{(1)}|\psi\rangle,
\end{equation}
where, for $j=1,2$, $\Pi_{n_j}^{(j)}$ 
projects onto the subspace 
$\mathfrak{S}\big[V_1^{\otimes n_1}\otimes V_2^{\otimes n_2}\big]$ which is isomorphic to $\mathfrak{h}_{n_1}\otimes\mathfrak{h}_{n_2}$ (see Section \ref{eff.dist}).
Then, $|\psi\rangle$ is
separable-III if $|\psi_{n_1,n_2}\rangle$ is separable as a state in the tensor product space $\mathfrak{h}_{n_1}\otimes\mathfrak{h}_{n_2}$ for all $n_{1,2}$~\cite{Wiseman2003}.

On the subspace $\mathfrak{S}\big[V_1^{\otimes n_1}\otimes V_2^{\otimes n_2}\big]\cong\mathfrak{h}_{n_1}\otimes\mathfrak{h}_{n_2}$ with fixed $n_j$, the elements of the subalgebras \eqref{subalg-part-I} act as
operators of the form $O_{n_1}\otimes\mathbbm{1}$ ($j=1$) or $\mathbbm{1}\otimes O_{n_2}$ ($j=2$).
Since
separability-III, when rephrased for the vectors in
$\mathfrak{h}_{n_1}\otimes\mathfrak{h}_{n_2}$, is formally equivalent to Definition \ref{part-sep}, the expectations of local operators in
the above subalgebras, when evaluated with respect to
separable-III states $|\psi_{n_1,n_2}\rangle$, do factorize.

The expectations of $(\mathcal{A}_1,\mathcal{A}_2)$-local operators, i.e. $O_1 O_2$ with $O_j\in\mathcal{A}_j$, with respect to a generic state $|\psi\rangle$ read
\begin{equation} \label{part-sep-C-exp}
\langle\psi|O_1O_2|\psi\rangle=\sum_{n_1,n_2=0}^\infty p_{n_1,n_2}\langle\psi_{n_1,n_2}|O_1 O_2|\psi_{n_1,n_2}\rangle.
\end{equation}
Each expectation in the sum factorize for any $O_1$ and $O_2$
if and only if $|\psi_{n_1,n_2}\rangle$ is
separable-III for all $n_1$ and $n_2$.
We therefore find
\begin{equation}
\langle\psi|O_1 O_2|\psi\rangle=\sum_{n_1,n_2=0}^\infty p_{n_1,n_2}\langle\psi_{n_1,n_2}|O_1|\psi_{n_1,n_2}\rangle \langle\psi_{n_1,n_2}|O_2|\psi_{n_1,n_2}\rangle
\label{part-sep-C-loc}
\end{equation}
which is of the form \eqref{sep2}. The fact that the expectations in \eqref{part-sep-C-loc} with respect to the given pure state are those of a density matrix with
weights $p_{n_1,n_2}\neq\delta_{n_1,\bar n_1}\delta_{n_2,\bar n_2}$ can be understood by observing that
the chosen subalgebras do not generate the entire operator algebra.
Despite this fact,
the above choice of commuting subalgebras is fully legitimate and shows, for any generic state $|\psi\rangle$, the agreement of
separability-III with
the notion of particle-locality presented in Definition~\ref{def-idloc}.
\footnote{The more general condition~\eqref{sep2} for mixed states originates from the convex roof construction~\cite{Horodecki2009}.}

\subsubsection{Effective distinguishability} \label{eff.dist.dist}

The notion of separability-III in Definition \ref{loc.ssr.ent} involves projecting pure states onto subspaces $\mathfrak{S}\big[V_j^{\otimes n_1}\otimes V_2^{\otimes n_2}\big]$. In Section \ref{eff.dist}, it is shown that the isomorphism between the Hilbert (sub)spaces $\mathfrak{S} \left[ V_1^{\otimes n_1} \otimes V_2^{\otimes n_2} \right]$ and $\mathfrak{h}_{n_1}\otimes\mathfrak{h}_{n_2}$ in Definition \ref{loc.ssr.ent}
is a possible way to implement effective distinguishability.

Indeed, the entanglement of two distinguishable particles is recovered from 
entanglement-III when identical particles can be effectively distinguished, e.g., by spatial localization.
As an example,
consider the setting consisting of particles described by an ``external'' degree of freedom with values $L,R$, and an ``internal'' degree of freedom $\sigma$, with values $\uparrow,\downarrow$.
Then,
one can naturally fix $V_1$ to be the subspace spanned by $L$-states,
$V_1=\textnormal{span}\{|L,\sigma\rangle\}_{\sigma}$,
and similarly $V_2$ by the $R$-states, $V_2=\textnormal{span}\{|R,\sigma\rangle\}_{\sigma}$,
with $n_1=n_2=1$. With this choice, the identification of separable states coincides with that for distinguishable particles. Also the evaluation of entanglement monotones provides results analogous to those for distinguishable particles.

Therefore, entanglement-III is fully consistent with the effective distinguishability of identical particles operated by freezing suitable number-degrees of freedom. 

\subsubsection{Information resources} \label{info.res.dist}

It has already been emphasized that
the projection onto orthogonal subspaces $V_j$, effectively embeds the standard entanglement of distinguishable
particles
into the formalism of identical particles. Therefore, the paradigm of local operators and classical communication \cite{Chitambar2014}, that of quantum non-locality \cite{BellGao}, and their applications
to quantum information \cite{NielsenChuang,BenattiFannesFloreaniniPetritis,Buhrman2010} can in this context be as well reformulated for  identical particles.

On the other hand, the projection operation in Definition \ref{loc.ssr.ent} causes a loss of quantum coherence, and therefore
entanglement-III is somewhat too restrictive as it cannot fully describe quantum effects that occur in informational tasks based on state vectors that live outside the special subspaces $\mathfrak{S} \left[ V_1^{\otimes n_1} \otimes V_2^{\otimes n_2} \right]$.

Using the notation introduced in Section \ref{info.res.perm}, a concrete example is provided by a general state of $N$ two-level identical particles \cite{Wiseman2004}
\begin{equation} \label{2lvl}
\sum_{k=0}^N C_k
\,\sqrt{\left(
\begin{matrix}
N \\
k
\end{matrix}
\right)} \,
\mathfrak{S}\big[|0\rangle^{\otimes k}\otimes|1\rangle^{\otimes (N-k)}\big],
\end{equation}
with $C_k\in\mathbbm{C}$ and $\sum_k|C_k|^2=1$.
The orthogonal subspaces as in Definition \ref{loc.ssr.ent} are $V_1=\{|0\rangle\}$ and $V_2=\{|1\rangle\}$, up to single-particle change of basis. The projection in Definition \ref{loc.ssr.ent}, with $n_1=k$ and $n_2=N-k$, reduces the above states to
\begin{equation} \label{perm-det}
\sqrt{\left(
\begin{matrix}
N \\
k
\end{matrix}
\right)} \,
\mathfrak{S}\big[|0\rangle^{\otimes k}\otimes|1\rangle^{\otimes (N-k)}\big]
\end{equation}
with probability $|C_k|^2$.
States as in~\eqref{perm-det} are
separable-III for they are
isomorphic to $|0\rangle^{\otimes k}\otimes|1\rangle^{\otimes (N-k)}\in\mathfrak{h}_k\otimes\mathfrak{h}_{N-k}$, according to equation \eqref{isomorphism} in~\ref{gen.isom}.
Thus, states in quantum protocols with two-level particles cannot, at any time during the process,
be entangled-III although these states provide quantum enhanced performances not only for interferometric phase estimation \cite{Benatti2010,Benatti2011,Argentieri2011,Benatti2013,
Benatti2014,Braun2018} but also for teleportation \cite{Schuch2004-1,Schuch2004-2,Heaney2009-2,Friis2013-2,
Marzolino2015,Marzolino2016}, entanglement distillation \cite{Schuch2004-1,Schuch2004-2}, quantum data hiding \cite{Verstraete2003}, and Bell's inequalities \cite{Summers1987-1,Summers1987-2,Summers1987-3,Halvorson2000,
Wildfeuer2007,Ashhab2007,Bancal2008,Ashhab2009,Heaney2009,
Heaney2010,Heaney2011,Friis2011,Brask2012}.
The same argument holds for three-level states.

Thus, entanglement-III is not adequate to account for the quantum-enhancing character of protocols that do not
leave the subspaces $\mathfrak{S}\left[ V_1^{\otimes n_1} \otimes V_2^{\otimes n_2}\right]$ invariant.
Indeed, the considerations on particle-locality developed in section \ref{loc.op.dist} imply that
entanglement-III consistently identifies non-local correlations only after effectively distinguishing particles through freezing suitably chosen number degrees of freedom.

\subsection{Entanglement-IV} \label{hybrid}

In this section we examine the approach to indistinguishable particle entanglement elaborated in~\cite{LoFranco2016,Chin2019}:
its explicit purpose is to treat indistinguishable particles by means of the first quantization formalism without particle labels being attached to single particle state vectors.
By its very construction, this approach  relies upon particles and single-particle operators; therefore,  the underlying locality criterion can be characterized
by Definition~\ref{def-idloc}.

In the case of two identical particles described by the (anti-)symmetrized Hilbert space $\mathfrak{S}\left[\mathcal{H}^{\otimes 2}\right]$, one introduces 
operators $A_\psi$ mapping the latter onto the single-particle Hilbert space $\mathcal{H}$, according to
\begin{equation} 
\label{Pi-psi}
A_{\psi} \, \mathfrak{S}\big[|\phi\rangle\otimes|\zeta\rangle\big]=\langle\psi|\phi\rangle|\zeta \, \rangle+\eta\langle\psi|\zeta\rangle|\phi \, \rangle\ ,
\end{equation}
where $\eta=+1$ ($-1$) for Bosons (Fermions), while the chosen set of single-particle vectors $\{|\psi_k\rangle\}_{k\in K}$ span a suitable subspace $\mathcal{K}\subseteq\mathcal{H}$.
These operators  are not projections and their adjoints $A^\dag_\psi:\mathcal{H}\mapsto\mathcal{S}\left[\mathcal{H}^{\otimes 2}\right]$ are 
\begin{equation} 
\label{Pi-psi-adj}
A^\dag_{\psi} \, \vert\phi\rangle=\mathfrak{S}\big[|\psi\rangle\otimes|\phi\rangle\big]\ .
\end{equation}
In the second quantization formalism (see~\ref{app.hybrid-ent}) $A_\psi$ and $A^\dag_\psi$ correspond to the annihilation and creation operators of the single-particle state 
$\vert\psi\rangle\in\mathcal{H}$; their very definition makes them insensitive to the particle label of the single-particle states.

\begin{definition}[Entanglement-IV] \label{hybrid-ent}
A pure state of two indistinguishable particles $\vert\psi\rangle\in\mathfrak{S}\left[\mathcal{H}^{\otimes 2}\right]$ is entangled relative to the subspace $\mathcal{K}\subseteq\mathcal{H}$, if the so-called "reduced single-particle density matrix'',  
\begin{equation}
 \label{1par-op.pa}
X_1:=\frac{1}{\displaystyle\sum_{k\in K}\|A^\dag_{\psi_k}\,\psi\|^2}\,\sum_{k\in K}A_{\psi_k}\,\vert\psi\rangle\langle\psi\vert \,A_{\psi_k}^\dag\ ,
\end{equation}
has non-vanishing  von Neumann entropy $S(X_1)=-\textnormal{tr}\big(X_1\log X_1\big)$, for any given choice of orthogonal $\vert\psi_k\rangle\in\mathcal{H}$, 
$k\in K$, spanning $\mathcal{K}$.
\end{definition}

The above definition and its properties have been presented and discussed in \cite{LoFranco2016,Sciara2017,Compagno2018,Chin2019,
Chin2019-2,Mani2020}
with specific applications to entanglement manipulation \cite{Bellomo2017}.

Evidently, the entanglement properties based on the above Definition depend on 
the subspace  $\mathcal{K}$; however, the orthonormality of the basis vectors $\vert\psi_k\rangle$ guarantees that $X_1$, and thus the entanglement properties, 
do not depend on the chosen basis in $\mathcal{K}$ (see~\ref{app.hybrid-ent}).
 
Notice that, as observed in Remark 3 of \cite{Benatti2017}, $X_1$ cannot be interpreted as a standard reduced density matrix in the one-particle sector; indeed, the expectation values of single-particles observables computed with respect to the mixed one-particle state $X_1$ do not reproduce those computed with respect to the original two-particle pure state $\vert\psi\rangle\in\mathfrak{S}\left[\mathcal{H}^{\otimes 2}\right]$.

Because of its very definition, entanglement-IV
is a new way of looking at the correlations between the degrees of freedom supported by $\mathcal{K}$ and those supported by its orthogonal complement. As
explicitly shown in~\ref{app.hybrid-ent}, in some cases entangled-IV states reduce to  entanglement-I ones, while in other cases  to entangled-III states, whence the notion of entanglement-IV  suffers from some of the problems of those two approaches with respect to the three discriminating entanglement criteria.

\subsubsection{Local operators} 

By construction, entanglement-IV is concerned with correlations in relation to a chosen single-particle subspace $\mathcal{K}$.
Given a Bosonic state $|\phi\rangle\otimes|\phi\rangle$, with $|\phi\rangle$ any single-particle state, its reduced density matrix relative to a generic single-particle subspace $\mathcal{K}$ is $\vert\phi\rangle\langle\phi\vert$. 
As a consequence, the state is separable-IV for whatever choice of $\mathcal{K}$; then, the argument raised in Section~\ref{loc.op.fact} applies thereby
showing that the locality criterion fails.
Indeed, for any choice of algebraic bipartition, there always exists
a separable-IV state $|\phi\rangle\otimes|\phi\rangle$
whose expectations on products of operators $A_i=\mathcal{P}(O_i,\mathbbm{1})$, $i=1,2$, belonging to the corresponding
commuting subalgebras, do not factorize.
Explicit examples in support to this conclusion are discussed in~\ref{app.loc.op.hybrid-ent}.

\subsubsection{Effective distinguishability} \label{eff.dist.fact_hybrid}

The notion of entanglement based on Definition~\ref{hybrid-ent} appears to be compatible with effective distinguishability.
As an example, let us consider again a two-particle system with spatial, $S=L,R$, and internal, $\sigma=\uparrow,\downarrow$, degrees of freedom. If $\mathcal{K}$ is 
the subspace spanned by 
$\vert L,\sigma\rangle$, $\sigma=\uparrow,\downarrow$, then
the only way to effectively distinguish the two particles is via the spatial labels.  Then, using the arguments and formalism of Section \ref{eff.dist}, one proves that all effectively spatially distinguished two-particle separable states  are also separable-IV. As discussed in Section \ref{eff.dist}, this argument can be generalized to any generic type of effective distinguishability.

\subsubsection{Information resources} \label{info.res.fact_hybrid}

The entangled-IV state $\mathfrak{S}\big[|L,\uparrow\rangle\otimes|L,\downarrow\rangle\big]$ was studied as a resource for probabilistic entanglement swapping,  an application of teleportation \cite{Castellini2019}. This protocol consists of either local measurements within the framework of entanglement-IV \cite{LoFranco2016} or operations that, although non-local, cannot change the amount of entanglement-IV quantified by the von Neumann entropy  $S(X_1)$.
In all these approaches, measurements of single-particle properties when
particles are spatially localized are considered local.

However, such local measurements, by acting
on separable-IV states like $|L,+\rangle\otimes|L,+\rangle$, with \hbox{$|+\rangle=\big(|\!\uparrow\rangle+|\!\downarrow\rangle\big)/\sqrt{2}$}, 
generate the entangled-IV state $\mathfrak{S}\big[|L,\uparrow\rangle\otimes|L,\downarrow\rangle\big]$.
Indeed, consider the separable-IV state
\begin{equation} \label{sep-nolabel}
|L,+\rangle\otimes|L,+\rangle, \qquad \textnormal{with} \qquad |+\rangle=\frac{|\uparrow\rangle+|\downarrow\rangle}{\sqrt{2}}\ .
\end{equation}
Measuring the observable $\mathcal{P}(P_L\otimes\sigma_z,\mathbbm{1})$, namely the spin along the $z$ axis of a single-particle localized at the left position, amounts to a local  operation according to \cite{LoFranco2016,Castellini2019}.
Nevertheless, among the (non-normalized) eigenvectors of $\mathcal{P}(P_L\otimes\sigma_z,\mathbbm{1})$,
\begin{equation}
\Big\{\mathfrak{S}\big[|S,\sigma\rangle\otimes|S',\sigma'\rangle\big]\Big\}_{S,S'=L,R;\ \sigma,\sigma'=\uparrow,\downarrow} \ ,
\end{equation}
two are entangled-IV states, namely $\mathfrak{S}\big[|S,\uparrow\rangle\otimes|S,\downarrow\rangle\big]$ with $S=L,R$.
Therefore, local measurement might transform the separable-IV state \eqref{sep-nolabel} into the entangled-IV state $\sqrt{2}\,\mathfrak{S}\big[|L,\uparrow\rangle\otimes|L,\downarrow\rangle\big]$ with probability $1/2$.

Consequently, an internal inconsistency arises between entanglement-IV and particle-local operations, whose locality
agrees with 
Definition~\ref{def-idloc}.
In addition, there is also  an operational paradox; indeed,
suppose the entanglement-IV  supported by the state 
$\mathfrak{S}\big[|L,\uparrow\rangle\otimes|L,\downarrow\rangle\big]$
is used  in the aforementioned entanglement swapping protocol. If it could be generated from $|L,+\rangle\otimes|L,+\rangle$, the overall protocol would perform an inherently quantum information task that however uses only local operations, classical communication and separable-IV resource states.

\bigskip

All four definitions of identical particle entanglement so far discussed make essentially use of the first 
quantization formalism, and thus are based on a locality criterion that focuses on correlations (or their lack thereof) among particles. However, in quantum statistical mechanics 
and quantum field theory the standard
approach to deal with many-body systems made of indistinguishable constituents  is provided by the second quantization approach.
In the next section, we will show how
full use of this formalism can lead to a fully physically consistent definition of  indistinguishable 
particle entanglement.

\section{Mode-partitioning}
\label{mode-ent}

In the previous section, we have analysed how entanglement
can be extended to 
identical particles in the way of particle-entanglement. 
In this section, we consider the  different perspective
given by second quantization
where one focuses no longer on the particle aspect, rather on generic degrees of freedom thus accommodating properties more general than those of particle type. 
As already presented in Section~\ref{eff.dist},  based on the formalism of 
annihilation and creation operators satisfying (anti-)commutation relations, these operators, acting on the Fock vacuum, annihilate, respectively create single-particle states,
also called \textit{modes}.

The general characterization of absence of correlations in~\eqref{sep2} can be still used when dealing with identical particles to define separable and entangled states. However, the lack of correlations will refer not to
observables related to single particle properties,
rather to subsets
of creation and annihilation operators identifying
groups of orthogonal modes that may also correspond to global properties;
these will be referred to as \textit{mode-partitions}.

More specifically, in the case of $M$ orthogonal modes, a bipartition of the set of creation, $\mathfrak{a}_j^\dagger$, and annihilation,
$\mathfrak{a}_j$, operators, $j=1, 2,\cdots, M$, is simply given by the two subsets:
\begin{equation}
\label{localg0}
\mathcal{A}_1=\{\mathfrak{a}_j,\mathfrak{a}_j^\dag\}_{j=1}^m\ ,\ \mathcal{A}_2=\{\mathfrak{a}_j,\mathfrak{a}_j^\dag\}_{j=m+1}^M\ ,
\end{equation}
\textit{i.e.} the sets of all polynomials in the corresponding annihilation and creation operators.
In
the second quantization approach local operators are products of elements of these two
subsets:
\begin{definition}[Mode-locality]
\label{mod-loc-def}
With respect to  mode-(bi)partitions, mode-local operators are of the form $A_1 A_2$ with $A_1\in\mathcal{A}_1$ and $A_2\in\mathcal{A}_2$.
\end{definition}

For simplicity, we have used a discrete set of modes, but the theory can be extended to continuous sets as, for instance,  in quantum optics when one considers processes that involve
photons with continuously varying wave-vectors.

Unlike particle-locality, the above kind of locality
embodies the request that measurements of mode-local operators in one
subset do not influence measurements of those in the other
subset. We shall refer to it as mode-locality, in relation to which one has then to distinguish between Bosons and Fermions. Indeed, because of the Canonical Commutation Relations (CCR), Bosonic mode-partitions of identical Bosons give rise to commuting
subsets when the modes are orthogonal to each other~\cite{Benatti2010}; for Fermions one needs to take into account the implications
of the Canonical Anticommutation Relations (CAR) \cite{Benatti2014,Benatti2016}.

In the Fermionic case there are two choices: either one complies with actual measurability of operators and thus restrict to the so-called even operators that are made of monomials with even number of annihilation and/or creation operators~\cite{Emch,Haag}, or with the request of the so-called \textit{microcausality}~\cite{StreaterWightman,Strocchi2004}; this last case asks for considering also odd operators but proves to be more suitable in most physical applications.
Fermions are also peculiar because they obey a parity superselection rule \cite{Wick1952,Aharonov1967,Hegerfeldt1968,Mirman1969,
Mirman1970,Wick1970,Strocchi1974,MullerHerold1985,Cisneros1998,
Bartlett2007} that prevents the superposition of a bosonic compound and a fermionic system.
These properties affects entanglement theory and its applications in quantum information
\cite{Moriya2005,Caban2005,Caban2005-2,Banuls2007,Banuls2009,
Bradler2011,Friis2013,Chiu2013,Friis2016}.

\medskip

\noindent
Using the above notion of mode-locality, the property \eqref{sep2} can now be used to identify a new definition of separability and entanglement for systems of identical particles, that will be denoted as \textit{separability-V} and \textit{entanglement-V}.

\begin{definition}[Entanglement-V]
\label{mode.ent.def}-
A generic mixed state is separable with respect to a mode-bipartition
$\mathcal{A}_1=\{\mathfrak{a}_j,\mathfrak{a}_j^\dag\}_{j=1,\dots,m}$ and $\mathcal{A}_2=\{\mathfrak{a}_j,\mathfrak{a}_j^\dag\}_{j=m+1,\dots,M}$ of a given 
set of orthogonal modes  if \eqref{sep2} holds for all $A_1\in\mathcal{A}_1$, $A_2\in\mathcal{A}_2$.
It then follows that a pure state is separable if and only if it can be written in the form 
\begin{equation}
\label{modsep}
P(\mathfrak{a}_1^\dag, \dots, \mathfrak{a}_m^\dag)\, Q(\mathfrak{a}_{m+1}^\dag, \dots, \mathfrak{a}_M^\dag) |\textnormal{vac}\rangle\ ,
\end{equation} 
by means of general polynomials $P$ and $Q$, where $|\textnormal{vac}\rangle$ is the Fock vacuum state.
Separable mixed states are convex combinations of separable pure states.
All other states are entangled.
\end{definition}
\medskip

As a specific application, let us consider the case of ``spatial'', $L,R$, and ``internal'', $\uparrow,\downarrow$, degrees of freedom introduced above,
and choose
the mode-bipartition
\begin{equation}
\label{bip}
\mathcal{A}_L=\{\mathfrak{a}_{L,\sigma},\mathfrak{a}_{L,\sigma}^\dag\}_{\sigma\in\{\uparrow,\downarrow\}}\ ,\
\mathcal{A}_R=\{\mathfrak{a}_{R,\sigma},\mathfrak{a}_{R,\sigma}^\dag\}_{\sigma\in\{\uparrow,\downarrow\}}\ .
\end{equation}
According to the above definition,
two-particle pure states
are
separable-V with respect to the bipartition 
$\mathcal{A}_{L}$, $\mathcal{A}_{R}$ if they are of the form

\begin{equation}
\sum_{\sigma,\sigma'}c_{\sigma}\, c_{\sigma'}\mathfrak{a}_{L,\sigma}^\dag\mathfrak{a}_{R,\sigma'}^\dag|\textnormal{vac}\rangle \ ,
\label{gen.eff.dist.state}
\end{equation}
as well as
\begin{equation}
\sum_{\sigma,\sigma'}c_{\sigma,\sigma'}\mathfrak{a}_{S,\sigma}^\dag\mathfrak{a}_{S,\sigma'}^\dag|\textnormal{vac}\rangle \ ,
\label{gen.eff.dist.state2}
\end{equation}
for any complex constants $c_\sigma$, $c_{\sigma,\sigma'}$, and $S$ being equal either to $L$ or to $R$.

Entanglement-V accounts for quantum correlations between occupancies of orthogonal modes. Indeed, unlike identical particles, orthogonal modes can always be individually addressed in experiments \cite{Wurtz2009,Bakr2010,Sherson2010}. 
Several fundamental properties of entanglement-V have been studied
\cite{Zanardi2001,Fang2003,Shi2003,Schuch2004-1,Schuch2004-2,
Caban2005,Caban2005-2,Cavalcanti2005,Banuls2007,Banuls2009,
Benatti2010,Bradler2011,Friis2013,Bose2013,Balachandran2013,
Balachandran2013-2,Benatti2014,Benatti2014-2,Benatti2014-3,
Ma2014,Karczewski2016,Friis2016,Benatti2016,Johansson2016,
Spee2018,Sperling2019,Ding2020},
as well as its experimental detection \cite{Bose2003,Anders2006,Ashhab2007-2,Goold2009,Lougovski2009,
Cramer2013,Dasenbrook2016,Fadel2018,Cornfeld2019}
and the possibility of practically manipulating it~\cite{Calsamiglia2002,Paunkovic2002,Omar2002,Bose2002,
Bose2002-2,Ray2011,Gagatsos2013,Marzolino2013,Islam2015,
Kaufman2018,Bernardo2018}.
Furthermore, entanglement-V can be quantified by means of {\it so-called} entanglement monotones, such as entanglement entropy
\cite{Zanardi2002-1,Gittings2002,Zanardi2003,
Gigena2015,Mondal2016,Gigena2017}, concurrence \cite{Zanardi2002-2,Caban2005,Caban2005-2}, geometric measures \cite{Lari2010}, negativity \cite{Benatti2012,Eisler2015,Shapourian2017,Eisert2018}, and robustness of entanglement \cite{Benatti2012-2}.
Entanglement-V has also been investigated in condensed matter systems
\cite{Zanardi2002-1,Zanardi2002-2,Zanardi2003,Hines2003,
Vedral2003,Shi2004,Giorda2004,Vedral2004,Li2008,Song2012,
Daley2012,Bruschi2013,Frerot2015,Frerot2016,Olsen2016,
Duvenhage2018,Barghathi2018,Barghathi2019}, quantum optics \cite{Kim2003,Papp2009,Matthews2009,Pan2012,Roslund2013,
Chen2014,Takeda2019,Fabre2019,Qian2018,Qian2020,Qian2020-2,
Sandbo2020}, and quantum field theories
\cite{Srednicki1993,Hawking2001,Clifton2001,Narnhofer2002,
Narnhofer2004,Casini2004,Calabrese2004,Narnhofer2005,
Verch2005,Calabrese2005,Ryu2006,Ryu2006-2,Pakman2008,
Casini2009,Calabrese2009,CastroAlvaredo2009,Casini2011,
Calabrese2012,Friis2012,Friis2012-2,Friis2012-3,Friis2013-3,
Friis2013-4,Bruschi2013,Coser2014,Yngvason2015,Cardy2016,
Calabrese2016,HollandSanders,Nishioka2018}.
However, entanglement-V is not an absolute property but depends on the choice of orthogonal modes \cite{Ciancio2006,Benatti2010,Benatti2012-2,Benatti2014}: mode transformations are mode non-local and thus do not preserve entanglement-V.

In addition, the above definitions generalize to mode-multipartitions. For instance, a multipartite pure state is fully separable with respect to a multipartition into each one of a set of $M$ orthogonal modes ($\{\mathfrak{a}_1,\mathfrak{a}_1^\dag\},\cdots,\{\mathfrak{a}_M,\mathfrak{a}_M^\dag\}$), if and only if it is of the form 
$\prod_{j=1}^M\frac{(\mathfrak{a}_j^\dag)^{k_j}}{\sqrt{k_j!}}\vert\textnormal{vac}\rangle$.

\subsection{Local operators} \label{loc.op-mode}

The structure of separable-V pure states in
Definition \ref{mode.ent.def}
is exactly derived from asking that the factorization as in~\eqref{sep2} over mode-local observables hold \cite{Benatti2010,Benatti2012,
Benatti2012-2,
Benatti2014,Benatti2014-2,Benatti2016}.
Furthermore, theory and applications of local operations and classical communication straightforwardly extend from the distinguishable particle case \cite{Chitambar2014} to the framework of entanglement-V: it is sufficient to replace distinguishable particle local operators, $O_1\otimes O_2$, by mode-local operators $O_1O_2$.
Also the Schmidt decomposition for pure states \cite{NielsenChuang,
BenattiFannesFloreaniniPetritis} can be generalized to the entanglement-V framework \cite{Hines2003,Schuch2004-1,Schuch2004-2,Benatti2012-2}.
Therefore, the non-increasing property of entanglement under local operations and classical communication follows from the same arguments valid for distinguishable particles.

\subsection{Effective distinguishability} \label{eff.dist.mode}

As previously stressed, distinguishability can be obtained by ``freezing'' some modes, for instance the spatial degrees
of freedom $L,R$ in the above considered case.
In the present context, separability of effectively distinguishable particles is then directly induced by the mode-bipartition $(\mathcal{A}_L,\mathcal{A}_R)$ in \eqref{bip}, as shown in the following (see also \cite{Moryia2006,Cunden2014}).

Considering the algebraic mode-bipartition $(\mathcal{A}_L,\mathcal{A}_R)$ in \eqref{bip}, the separable-V state 
\begin{equation}
\label{auxx}
\left(\sum_\sigma c^{(1)}_\sigma\mathfrak{a}_{L,\sigma}^\dag\right)\left(\sum_{\sigma'}c^{(2)}_{\sigma'}\mathfrak{a}_{R,\sigma'}^\dag\right)|\textnormal{vac}\rangle
\ ,
\end{equation}
with $c^{(1,2)}_\sigma$ complex numbers, written by means of the notation introduced in~\eqref{auxx2}, corresponds in first quantization to the standard separable two-particle state

\begin{eqnarray}
\label{auxx1}
&&
\vert\psi_L\rangle\otimes\vert\psi_R\rangle\ , 
\qquad\hbox{where}\\
\nonumber
&&
\vert\psi_{L}\rangle=\sum_\sigma c^{(1,2)}_{\sigma}\vert L,\sigma\rangle\ ,\
\vert\psi_{R}\rangle=\sum_\sigma c^{(1,2)}_{\sigma}\vert R,\sigma\rangle\ .
\end{eqnarray}
In other terms, separability-V reduces to standard separability in the case of systems of distinguishable particles (see Definition \ref{part-sep}).

Then, entanglement-V monotones relative to the mode-partition \hbox{$(\mathcal{A}_L,\mathcal{A}_R)$}, such as entanglement entropy \cite{Zanardi2002-1,Gittings2002,Zanardi2003}, concurrence \cite{Zanardi2002-2,Caban2005}, geometric measures \cite{Lari2010}, negativity \cite{Benatti2012,Eisler2015,Shapourian2017,Eisert2018}, and robustness of entanglement \cite{Benatti2012-2}, also reduce to standard expressions through effective distinguishability.

Let us remark that the operators $A_L$ and $A_R$ in equations (\ref{AL},\ref{AR}) are special instances of the operators belonging to the subalgebras $\mathcal{A}_L$ and $\mathcal{A}_R$;
indeed, they commute with the particle numbers 
\begin{equation} \label{part-numb-LR}
N_L:=\sum_{\sigma}\mathfrak{a}^\dag_{L,\sigma}\mathfrak{a}_{L,\sigma}\ ,\quad N_R:=\sum_{\sigma}\mathfrak{a}^\dag_{R,\sigma}\mathfrak{a}_{R,\sigma}\ .
\end{equation}
If these commutation properties are enforced on the algebraic bipartition, the entanglement-V approach recovers entanglement-III. In general, operators in $\mathcal{A}_L$ and $\mathcal{A}_R$ do not commute with $N_{L,R}$ and thus change the particle number content: for instance, $\mathfrak{a}_{L,\sigma}^\dag$ or $\mathfrak{a}_{R,\sigma'}^\dag$  map sectors of the Fock space with $N_{R,L}$ particles into sectors with $N_{L,R}+1$ particles. These operators, which make perfect sense in a micro-causal approach to mode-locality, cannot be represented in the standard formalism of first quantization where the particle number is conserved. For instance, the state in equation \eqref{ex} can be created from the vacuum with local operators in the algebraic bipartition $(\mathcal{A}_L,\mathcal{A}_R)$, and is thus separable; however, it cannot effectively correspond to any state of distinguishable particles. In this respect the mode-bipartition $(\mathcal{A}_L,\mathcal{A}_R)$ not only covers the notion of entanglement for distinguishable particles, but also extends it to the whole Fock space.

\subsection{Information resources} \label{info.res.mode}

The consistent use of entanglement-V as a resource for quantum enhancements in information tasks has been already shown in several protocols.
Teleportation \cite{Schuch2004-1,Schuch2004-2,Heaney2009-2,Friis2013-2,
Marzolino2015,Marzolino2016}, entanglement distillation \cite{Schuch2004-1,Schuch2004-2}, quantum data hiding \cite{Verstraete2003}, and practical checks of Bell's inequalities \cite{Summers1987-1,Summers1987-2,Summers1987-3,Halvorson2000,
Wildfeuer2007,Ashhab2007,Bancal2008,Ashhab2009,Heaney2009,
Heaney2010,Heaney2011,Friis2011,Brask2012} are information protocols consisting of mode-local operations and classical communication, where the use of entangled-V states leads to quantum enhanced performances.
The same holds for the precision estimation of interferometric phases \cite{Benatti2010,Benatti2011,Argentieri2011,Benatti2013,
Benatti2014,Braun2018}, which also provides examples of non-local protocols, namely mode-nonlocal interferometers, which generate the entanglement-V needed to overcome classical performances based on separable-V initial states.

A more general consistency check relies upon the compatibility, mentioned in Section~\ref{loc.op-mode}, of entanglement-V with classical communication and operations that are local with respect to a chosen mode-partition.
This property, together with the aforementioned concrete information protocols, endows entanglement-V with the complete operational interpretation of a resource theory \cite{NielsenChuang,BenattiFannesFloreaniniPetritis,Buhrman2010}, and yet with different practical consequences if compared with the case of distinguishable particles \cite{Verstraete2003,Marzolino2015,Marzolino2016}.

\section{Outlook} \label{conclusion}

We have considered different possible approaches to the definition of entanglement 
for quantum systems made of indistinguishable particles
and tested them against three natural physical consistency requests: \textit{i)}
entanglement should emerge as a manifestation of non-local correlations between commuting subalgebras of operators;
\textit{ii)} the entanglement of effectively distinguished identical particles should agree with the standard distinguishable particle entanglement; \textit{iii)} in absence of other quantum resources, entanglement should be identifiable as the only means to overcome classical performances.

Within the setting provided by the previous criteria, the outcome of
the test is that all formulations
considering entanglement as a property of particles 
do not comply with at least one of them, as summarized in the Table~\ref{table}.
On the other hand, entanglement-V, describing quantum correlations among second-quantized modes,
conforms to all three criteria, thus providing an adequate notion of identical particles entanglement with respect to the chosen setting.

In summary, we have analysed the variety of approaches to the notion of entanglement for systems made of indistinguishable particles. As a guiding property, we have focused on non-locality expressed as the lack of factorization of correlation functions involving commuting observables. This framework can be adopted in full generality both for distinguishable and indistinguishable particles, giving rise to the standard and agreed-upon notion of entanglement in the case of distinguishable particles. We hope that the results of our investigation may contribute to the debate on identical particle entanglement, with the aim at arriving to a fully consistent formulation which is all the more necessary in view of the ever more important impact of many-body systems in quantum technological applications.

\begin{table}[htbp]
\center{
\begin{tabular}{|c|c|c|c|}
\hline
& \multicolumn{3}{|c|}{compatibility criteria for entanglement definitions\phantom{\Big|}} \\
\hline
entanglement definition & local operators & effective distinguishability & information resources\phantom{\Big|} \\
\hline
\hline
entanglement-I & \xmark & \xmark & {\bf ?}\phantom{\Big|} \\
\hline
entanglement-II & \xmark & \xmark & \xmark\phantom{\Big|} \\
\hline
entanglement-III & \cmark & \cmark & \xmark\phantom{\Big|} \\
\hline
entanglement-IV & \xmark & \cmark & \xmark\phantom{\Big|} \\
\hline
entanglement-V & \cmark & \cmark & \cmark\phantom{\Big|} \\
\hline
\end{tabular}
}
\caption{\small{Summary of the consistency test: \cmark {} means that the entanglement definition complies with the criterion, \xmark {} means that it does not, while {\bf ?} means that, at the best of our knowledge, there is no evidence that the criterion is not violated.}
}
\label{table}
\end{table}

\appendix

\section{Identical particle effective distinguishability} \label{gen.isom}

The correspondence between identical and effectively distinguishable particles, as discussed in Section \ref{eff.dist} is an instance of a general isomorphism between suitable Fock subspaces
and tensor product Hilbert spaces. In this Section, we sketch how to
realize this isomorphism
with $N\geq 3$ identical  particles.
Let  $\Pi_\pi$ be the operator implementing the permutation $\pi:(1,2,\ldots,N)\to(\pi(1),\pi(2),\ldots,\pi(N))$ of particle labels,
so that
\begin{equation} \label{perm.op}
\Pi_\pi\bigotimes_{j=1}^N|\psi_j\rangle=\bigotimes_{j=1}^N|\psi_{\pi(j)}\rangle\ ,
\end{equation}
for any set of single-particle states $\{|\psi_j\rangle\}_{j=1,\cdots,N}$.
Then, the symmetrization, $\mathcal{S}$, and anti-symmetrization, $\mathcal{A}$, operators read
\begin{equation}
\mathfrak{S}=\frac{1}{N!}\sum_\pi x_\pi\Pi_\pi\ ,
\end{equation}ar
with
$x_\pi=1$ when $\mathfrak{S}=\mathcal{S}$ and $x_\pi=(-1)^{p(\pi)}$ when
$\mathfrak{S}=\mathcal{A}$, 
where $p(\pi)$ is the parity of the permutation $\pi$. An operator $O$ is permutation-invariant  if it commutes with all $\Pi_\pi$, namely if  
$O=\sum_\pi \Pi_\pi O\Pi_\pi$, as entailed by any superselection rule~\cite{Bartlett2007}. Therefore, to any particle-local observable of $N$ distinguishable particles 
$\bigotimes_{j=1}^N O_j$, one associates the following permutation-invariant generalization of equation \eqref{idensingop}:
\begin{equation} \label{PO1N}
\mathcal{P}(O_1,\cdots,O_N)=\sum_\pi\Pi_\pi\,\Big(O_1\otimes O_2\cdots\otimes O_N\Big)\,\Pi_\pi=\sum_\pi\bigotimes_{j=1}^N O_{\pi(j)}\ .
\end{equation}
One can thus construct
an isomorphism that extends the one between the states in equations (\ref{eff.dist.sep},\ref{eff.dist.ent})
and the states in equations (\ref{eff.dist.sep.isom},\ref{eff.dist.ent.isom}), or
those  in equation \eqref{gen.eff.dist.state}, to general $N$ identical particle pure states. 

Assume the single-particle Hilbert space to be a tensor product $\mathcal{H}=\mathcal{H}_{ext}\otimes\mathcal{H}_{int}$ of  Hilbert spaces associated to an external and internal degrees of freedom.
Let $\{\vert\psi_j\rangle\}_{j=1}^N$ be orthonormal vectors in $\mathcal{H}_{ext}$ and set $P_j=\vert\psi_j\rangle\langle\psi_j\vert$, with $P_jP_k=\delta_{jk}\,P_j$.
The symmetrized projector  $\mathcal{P}(P_1\otimes\mathbbm{1},\cdots,P_N\otimes\mathbbm{1})$ can be used to effectively distinguish identical particles. Indeed, 
one can show~\cite{Herbut1987,Herbut2001,Herbut2006} that  the subspace 
$$
\mathcal{P}(P_1\otimes\mathbbm{1},\cdots,P_N\otimes\mathbbm{1})\mathfrak{S}\big[\bigotimes_{j=1}^N\mathcal{H}\big]
$$ 
is mapped unitarily onto
the Hilbert space $\bigotimes_{j=1}^N\mathcal{H}_{int}$ by the operator $U$ such that
\begin{equation}
\label{isomorphism}
U\left(\sqrt{N!} \, \mathfrak{S} \bigotimes_{j=1}^N\vert\psi_j,\sigma_j\rangle\right)=
\bigotimes_{j=1}^N\vert\sigma_j\rangle\ .
\end{equation}

Clearly, in order to preserve effective distinguishability, one has to limit the accessible 
observables to those that commute with the
projector $\mathcal{P}(P_1\otimes\mathbbm{1},\cdots,P_N\otimes\mathbbm{1})$,
as $\mathcal{P}(P_1\otimes\Sigma_1,\cdots,P_N\otimes\Sigma_N)$.
Under the isomorphism \eqref{isomorphism}, symmetrized operators 
$\mathcal{P}(P_1\otimes\Sigma_1,\cdots,P_N\otimes\Sigma_N)$
behave as $\bigotimes_{j=1}^N\Sigma_j$
(compare with the expectations in equations \eqref{effdist} and \eqref{effdist2}).

\section{Entanglement-IV} \label{app.hybrid-ent}

As an example of separable-IV and entangled-IV states, let us consider the single-particle Hilbert space $\mathbb{C}^4$ spanned by the left-, right-localized vectors with spin $\sigma$, $\{\vert S,\sigma\rangle\}_{S=L,R;\sigma=\uparrow,\downarrow}$,  and the two-dimensional subspace  $\mathcal{K}$ spanned by spatially left-localized single-particle states $\textnormal\{|L,\sigma\rangle\}_{\sigma=\uparrow,\downarrow}$; namely, we will consider the entanglement relative to the internal, say spin, degrees of freedom.
For two indistinguishable particles within a same spatial region in a  state $\vert\psi\rangle=\sqrt{2}\,\mathfrak{S}\big[|L,\uparrow\rangle\otimes|L,\downarrow\rangle\big]$, entanglement-IV coincides with entanglement-I, where such a state is surely spin-entangled. 
Indeed, according to~\eqref{Pi-psi}, 
\begin{align}
A_{L\uparrow}\mathfrak{S}\big[|L,\uparrow\rangle\otimes|L,\downarrow\rangle\big]&=\vert L,\downarrow\rangle\\ 
A_{L\downarrow}\mathfrak{S}\big[|L,\uparrow\rangle\otimes|L,\downarrow\rangle\big]&=\vert L,\uparrow\rangle\ ,
\end{align}
so that 
\begin{equation}
\label{hybrid-aux0}
X_1=\frac{1}{2} \vert L,\downarrow\rangle\langle L,\downarrow\vert+\frac{1}{2} \vert L,\uparrow\rangle\langle L,\uparrow\vert\ ,
\end{equation}
whence $S(X_1)=\log 2$ and $\vert\psi\rangle$ is entangled with respect to the chosen subspace.
Instead, the vector state
$\vert\psi\rangle=|L,\uparrow\rangle\otimes|L,\uparrow\rangle$ which is separable-I is also separable-IV; indeed,  
\hbox{$X_1=| L,\uparrow\rangle\langle L,\uparrow\!|$} whence  $S(X_1)=0$.

On the other hand, for particles localized in non-overlapping spatial regions, entanglement-IV with respect to the subspace $\mathcal{K}$ chosen above coincides with entanglement-III.
Indeed, modifications of Definition \ref{hybrid-ent} restricted to the just mentioned physical situation has been considered in \cite{LoFranco2018,Barros2019}.
For instance, states of the form $\vert\Psi\rangle=\mathfrak{S}\big[\sum_{\sigma,\sigma'}c_{\sigma,\sigma'}|L,\sigma\rangle\otimes|R,\sigma'\rangle\big]$ are entangled-IV; indeed, one computes
\begin{equation}
\label{entIVex}
X_1=\mu_\uparrow\vert\psi_{R\uparrow}\rangle\langle\psi_{R\uparrow}\vert\,+\,\mu_\downarrow\vert\psi_{R\downarrow}\rangle\langle\psi_{R\downarrow}\vert\ ,ar
\end{equation}
where, with $\tau=\uparrow,\downarrow$,
$$
\vert\psi_{R\tau}\rangle:=\mu_\tau^{-1}\sum_{\sigma=\uparrow,\downarrow}
c_{\tau,\sigma}\vert R,\sigma\rangle\ ,\quad \mu_\tau:=\sum_{\sigma=\uparrow,\downarrow}\left|c_{\tau,\sigma}\right|^2\ .
$$
It follows that $S(X_1)=-\mu_\uparrow\log\mu_\uparrow-\mu_\downarrow\log\mu_\downarrow$ and then $\vert\Psi\rangle$ entangled-IV with respect to the chosen subspace $\mathcal{K}$, unless the coefficients factorize, $c_{\sigma,\sigma'}=c_\sigma\, c_{\sigma'}$. The latter condition also discriminates between $\vert\Psi\rangle$ being a separable-III or an entangled-III state. This result stems from Definition~\ref{loc.ssr.ent} by choosing $n_1=n_2=1$ and projecting them onto the state with each particle supported by one of the subspaces $V_1=\textnormal{span}\{|L,\sigma\rangle\}_{\sigma}$, $V_2=\textnormal{span}\{|R,\sigma\rangle\}_{\sigma}$.

With the second-quantization notation of Section~\ref{mode-ent}, the operators $A_\psi$  in~\eqref{Pi-psi} and $A^\dag_\psi$ in~\eqref{Pi-psi-adj} correspond to the annihilation and creation operators
$\mathfrak{a}_\psi$ and $\mathfrak{a}^\dag_\psi$ and the state $\mathfrak{S}\big[|\phi\rangle\otimes|\zeta\rangle\big]$ to $\mathfrak{a}^\dag_\phi\mathfrak{a}^\dag_\zeta\vert\textnormal{vac}\rangle$.
This fact can be directly derived by writing the annihilation and creation operators
of single-particle states in terms of the set $\{\mathfrak{a}_j,\mathfrak{a}^\dag_j\}_j$ relative to the single-particle orthonormal basis 
$\{\vert j\rangle\}_j$, as 
\begin{equation}
\label{hybrid-loc0a}
\mathfrak{a}_\psi=\sum_{j}\psi_j^*\,\mathfrak{a}_j\ ,\qquad \mathfrak{a}^\dag_\psi=\sum_{j}\psi_j\,\mathfrak{a}^\dag_j\ ,
\end{equation}
where $\vert\psi\rangle=\sum_j\psi_j\,\vert j\rangle$, $\vert j\rangle=\mathfrak{a}_j^\dag\vert\textnormal{vac}\rangle$, and by applying the ensuing (anti-)commutation relation
\begin{equation}
\label{hybrid-loc0b}
\mathfrak{a}_\psi\,\mathfrak{a}^\dag_\phi\,+\,\eta\,\mathfrak{a}^\dag_\phi\,\mathfrak{a}_\psi=\langle\psi\vert\phi\rangle\ ,\qquad \eta=\pm 1\ .
\end{equation}
Using $\mathfrak{a}_\psi$ and $\mathfrak{a}_\psi^\dag$, the one-particle density matrix in~\eqref{1par-op.pa} can be expressed as the result of the action of the following selective quantum channel
\begin{equation}
\label{red-mat}
X_1=\frac{1}{\displaystyle\sum_{k\in K}\|\mathfrak{a}_{\psi_k}\,\psi\|^2}\,\sum_{k\in K}
\mathfrak{a}_{\psi_k}\,\vert\psi\rangle\langle\psi\vert\,\mathfrak{a}_{\psi_k}^\dag\ .
\end{equation}
Changing orthonormal basis in $\mathcal{K}$ does not change $X_1$; indeed, sending $\vert\psi_k\rangle\mapsto\vert\phi_\ell\rangle=\sum_{j\in K}U_{\ell j}\vert\psi_k\rangle$ by means of a unitary matrix $U=[U_{\ell j}]$ yields
$\mathfrak{a}^\dag_{\phi_\ell}= \sum_{j\in K}U_{\ell j}\mathfrak{a}^\dag_{\psi_j}$, whence
\begin{equation}
\label{hybrid-loc0c}
\sum_{k\in K}\mathfrak{a}_{\psi_k}\,\vert\psi\rangle\langle\psi\vert\,\mathfrak{a}_{\psi_k}^\dag=\sum_{\ell\in K}\mathfrak{a}_{\phi_\ell}\,\vert\psi\rangle\langle\psi\vert\,\mathfrak{a}_{\phi_\ell}^\dag\ .
\end{equation}

\subsection{Local operators} \label{app.loc.op.hybrid-ent}

The argument in Section \ref{loc.op.fact} shows that, for any algebraic bipartition generated by single-particle operators, there is a separable-I state that does not satisfy the factorization of local expectation values \eqref{auxxx30}. Since pure separable-I states $|\phi\rangle\otimes|\phi\rangle$ are also separable-IV, the locality condition fails also in the Definition \ref{hybrid-ent}.

As a specific example, consider the bipartition generated by single-particle operators $A_1=\mathcal{P}\big(\vert L,\alpha\rangle\langle L,\alpha\vert,\mathbbm{1}\big)$ and $A_2=\mathcal{P}\big(\vert L,\alpha^\perp\rangle\langle L,\alpha^\perp\vert,\mathbbm{1}\big)$, with $|\alpha\rangle$ and $|\alpha^\perp\rangle$ two orthogonal states
relative to the single-particle internal degree of freedom. The separable-IV state $|L,\uparrow\rangle\otimes|L,\uparrow\rangle$ satisfies the factorization condition
\begin{align}
\label{hybrid-loc}
& \langle\psi\vert A_1\,A_2\vert\psi\rangle=2\big|\langle\alpha|\uparrow\rangle\langle\alpha^\perp|\uparrow\rangle\big|^2 \nonumber \\
& =\langle\psi\vert A_1\vert\psi\rangle\langle\psi\vert A_2\vert\psi\rangle
=4\big|\langle\alpha|\uparrow\rangle\langle\alpha^\perp|\uparrow\rangle\big|^2 \ .
\end{align}
if only if $\alpha=\uparrow$ \cite{Benatti2017}. On the other hand, the separable-IV state $|L,+\rangle\otimes|L,+\rangle$, with \mbox{$|+\rangle=\big(|\!\uparrow\rangle+|\!\downarrow\rangle\big)/\sqrt{2}$} satisfies the factorization condition if only if $\alpha=+$. Therefore, the two aforementioned states cannot be simultaneously separable.\footnote{To obtain a more general and concise treatment, one can work within the second-quantized formalism.
Consider a class of bipartitions by selecting  orthogonal subspaces $\mathcal{K}_{1,2}$ of $\mathcal{K}$ 
Then, the separable-IV state 
$\vert\psi_{12}\rangle=\vert\phi\rangle\otimes\vert\phi\rangle$ corresponds to $\vert\psi_{12}\rangle=\big((\mathfrak{a}_\phi^\dag)^2/{\sqrt{2}}\big)\vert \textnormal{vac}\rangle$, while 
single-particle operators of the form $A_{1,2}=\mathcal{P}\big(O_{1,2},\mathbbm{1}\big)$ where $O_i$ 
does not vanish on $\mathcal{K}_i$ only, are recast as
$A_{1,2}=2\sum_{k=1}^{K_{1,2}}o^{1,2}_k\,\mathfrak{a}^\dag_{\psi^{(1,2)}_k}\mathfrak{a}_{\psi^{(1,2)}_k}$,
where the orthonormal vectors $\{\vert\psi^{(1)}_k\rangle\}_{k=1}^{K_1}$ and  $\{\vert\psi^{(2)}_k\rangle\}_{k=1}^{K_2}$are orthonormal bases in $\mathcal{K}_{1,2}$, and $o^{1,2}_k$ the associated eigenvalues of $O_{1,2}$.
Then one finds again a contradiction:
$\langle\psi_{12}\vert A_1\,A_2\vert\psi_{12}\rangle\neq\langle\psi_{12}\vert A_1\vert\psi_{12}\rangle\langle\psi_{12}\vert A_2\vert\psi_{12}\rangle$.
}
Consider, instead, the bipartition generated by spatially localized single-particle operators, $A_1=\mathcal{P}\big(\vert L\rangle\langle L\vert\otimes O_1,\mathbbm{1}\big)$ and $A_2=\mathcal{P}\big(\vert R\rangle\langle R\vert\otimes O_2,\mathbbm{1}\big)$ for arbitrary operators $O_{1,2}$ of internal degrees of freedom, and separable-IV pure states
\begin{equation}
|\psi\rangle=\left(\frac{|L,\alpha\rangle+|R,\alpha\rangle}{\sqrt{2}}\right)^{\otimes 2} \ ,
\end{equation}
with $|\alpha\rangle$ an arbitrary state of the single-particle internal degree of freedom.
The factorizaton condition is not satisfied in general:
\begin{align}
& \langle\psi\vert A_1\,A_2\vert\psi\rangle=\frac{1}{2}\langle\alpha|O_1|\alpha\rangle\langle\alpha|O_2|\alpha\rangle \nonumber \\
& \neq\langle\psi\vert A_1\vert\psi\rangle\langle\psi\vert A_2\vert\psi\rangle
=\langle\alpha|O_1|\alpha\rangle\langle\alpha|O_2|\alpha\rangle \ .
\end{align}

{\bf Acknowledgments} F.F. and U.M. acknowledge support from the European Union through the H2020 CSA Twinning project No. 692194, ``RBI-T-WINNING'', and through the European Regional Development Fund - the Competitiveness and Cohesion Operational Programme (KK.01.1.1.06). U.M. is financially supported by the European Union's Horizon 2020 research and innovation programme under the Marie Sk\l odowska-Curie grant agreement No. 754496 - FELLINI. F.F. is partially supported by the Croatian Science Fund Project No. IP-2016-6-3347. F.B, R.F. and U.M. acknowledge that their research has been conducted within the framework of the Trieste Institute for Theoretical Quantum Technologies (TQT).

\bibliographystyle{naturemag}

\addcontentsline{toc}{section}{References}
\bibliography{ent_indist}

\end{document}